%% file: main.tex
\documentclass[twocolumn]{aastex63}
\usepackage{amsmath}
\usepackage{longtable}

\shorttitle{V1298 Tau RVs Using MAROON-X}
\shortauthors{Sikora et al.}

\graphicspath{{./}{figures/}}

\begin{document}

\title{Updated Planetary Mass Constraints of the Young V1298 Tau System Using MAROON-X}

\correspondingauthor{James Sikora}
\email{james.t.sikora@gmail.com}

\author[0000-0002-3522-5846]{James Sikora}
\affiliation{Anton Pannekoek Institute for Astronomy, University of Amsterdam, 1098 XH Amsterdam, The Netherlands}
\affiliation{Department of Physics \& Astronomy, Bishop's University, 2600 Rue College, Sherbrooke, QC J1M 1Z7, Canada}

\author[0000-0002-5904-1865]{Jason Rowe}
\affiliation{Department of Physics \& Astronomy, Bishop's University, 2600 Rue College, Sherbrooke, QC J1M 1Z7, Canada}

\author[0000-0000-0000-0000]{Saugata Barat}
\affiliation{Anton Pannekoek Institute for Astronomy, University of Amsterdam, 1098 XH Amsterdam, The Netherlands}

\author[0000-0003-4733-6532]{Jacob L.\ Bean}
\affiliation{Department of Astronomy \& Astrophysics, University of Chicago, Chicago, IL 60637, USA}

\author[0000-0003-2404-2427]{Madison Brady}
\affiliation{Department of Astronomy \& Astrophysics, University of Chicago, Chicago, IL 60637, USA}

\author[0000-0000-0000-0000]{Jean-Michel D\'esert}
\affiliation{Anton Pannekoek Institute for Astronomy, University of Amsterdam, 1098 XH Amsterdam, The Netherlands}

\author[0000-0002-9464-8101]{Adina D.\ Feinstein}
\altaffiliation{NSF Graduate Research Fellow}
\affiliation{Department of Astronomy \& Astrophysics, University of Chicago, Chicago, IL 60637, USA}

\author[0000-0002-0388-8004]{Emily A. Gilbert}
\affiliation{Jet Propulsion Laboratory, California Institute of Technology, 4800 Oak Grove Drive, Pasadena, CA 91109, USA}

\author[0000-0000-0000-0000]{Gregory Henry}
\affiliation{Center of Excellence in Information Systems, Tennessee State University, Nashville, TN 37209 USA}

\author[0000-0003-0534-6388]{David Kasper}
\affiliation{Department of Astronomy \& Astrophysics, University of Chicago, Chicago, IL 60637, USA}

\author[0000-0000-0000-0000]{D\'ereck-Alexandre Lizotte}
\affiliation{Department of Physics \& Astronomy, Bishop's University, 2600 Rue College, Sherbrooke, QC J1M 1Z7, Canada}

\author[0000-0002-1119-7473]{Michael R. B. Matesic}
\affiliation{D{\'e}partement de Physique, Universit{\'e} de Montr{\'e}al, 2900 Bd Édouard-Montpetit, Montr{\'e}al, QC H3T 1J4, Canada}
\affiliation{Department of Physics \& Astronomy, Bishop's University, 2600 Rue College, Sherbrooke, QC J1M 1Z7, Canada}
\affiliation{Trottier Institute for Research on Exoplanets, D{\'e}partement de Physique, Universit{\'e} de Montr{\'e}al, 2900 Bd Édouard-Montpetit, Montr{\'e}al, QC H3T 1J4, Canada}

\author[0000-0002-2513-4465]{Vatsal Panwar}
\affiliation{Department of Physics, University of Warwick, Coventry, West Midlands, CV47AL, UK}
\affiliation{Anton Pannekoek Institute for Astronomy, University of Amsterdam, 1098 XH Amsterdam, The Netherlands}

\author[0000-0003-4526-3747]{Andreas Seifahrt}
\affiliation{Department of Astronomy \& Astrophysics, University of Chicago, Chicago, IL 60637, USA}

\author[0000-0000-0000-0000]{Hinna Shivkumar}
\affiliation{Anton Pannekoek Institute for Astronomy, University of Amsterdam, 1098 XH Amsterdam, The Netherlands}

\author[0000-0001-7409-5688]{Gudmundur Stef\'ansson}
\affiliation{Department of Astrophysical Sciences, Princeton University, 4 Ivy Lane, Princeton, NJ 08540, USA}
\affiliation{NASA Sagan Fellow}

\author[0000-0002-4410-4712]{Julian St{\"u}rmer}
\affiliation{Landessternwarte, Zentrum f{\"u}r Astronomie der Universität Heidelberg, K{\"o}nigstuhl 12, D-69117 Heidelberg, Germany}

\begin{abstract}
The early K-type T-Tauri star, V1298 Tau ($V=10\,{\rm mag}$, ${\rm age}\approx20-30\,{\rm Myr}$) hosts four transiting planets with radii ranging from $4.9-9.6\,R_\oplus$. The three inner planets have orbital periods of $\approx8-24\,{\rm d}$ while the outer planet's period is poorly constrained by single transits observed with \emph{K2} and \emph{TESS}. Planets b, c, and d are proto-sub-Neptunes that may be undergoing significant mass loss. Depending on the stellar activity and planet masses, they are expected to evolve into super-Earths/sub-Neptunes that bound the radius valley. Here we present results of a joint transit and radial velocity (RV) modelling analysis, which includes recently obtained \emph{TESS} photometry and MAROON-X RV measurements. Assuming circular orbits, we obtain a low-significance ($\approx2\sigma$) RV detection of planet c implying a mass of $19.8_{-8.9}^{+9.3}\,M_\oplus$ and a conservative $2\sigma$ upper limit of $<39\,M_\oplus$. For planets b and d, we derive $2\sigma$ upper limits of $M_{\rm b}<159\,M_\oplus$ and $M_{\rm d}<41\,M_\oplus$. For planet e, plausible discrete periods of $P_{\rm e}>55.4\,{\rm d}$ are ruled out at a $3\sigma$ level while seven solutions with $43.3<P_{\rm e}/{\rm d}<55.4$ are consistent with the most probable $46.768131\pm000076\,{\rm d}$ solution within $3\sigma$. Adopting the most probable solution yields a $2.6\sigma$ RV detection with mass a of $0.66\pm0.26\,M_{\rm Jup}$. Comparing the updated mass and radius constraints with planetary evolution and interior structure models shows that planets b, d, and e are consistent with predictions for young gas-rich planets and that planet c is consistent with having a water-rich core with a substantial ($\sim5\%$ by mass) H$_2$ envelope.
\end{abstract}


\section{Introduction}\label{sect:intro}
 
Planets orbiting young stars ($\lesssim 100$\,Myr) serve as important windows into the early stages of planet formation and evolution. When coupled with known ages and insolation fluxes, bulk density measurements of young planets can be used to infer the core compositions and masses of their primordial H/He-dominated atmospheres \citep{fortney2007, lopez2014}. Additionally, precise mass constraints of such planets provide the unique opportunity to test, inform, and constrain initial planet formation location theories \citep{lee2015, lee2016, owen2020b} and theories of atmospheric mass loss processes \citep{kulow2014,oklopcic2018}. Transiting young planets are notoriously challenging to detect given the dominating underlying stellar variability of young stars. From the primary \emph{Kepler} mission \citep{borucki2010a}, it was uncovered that young planets are relatively rare \citep[$\lesssim4\%$ of transiting planets discovered have ages $<1\,{\rm Gyr}$,][]{berger2020b}. Since the launch of the Transiting Exoplanet Survey Satellite \citep[TESS;][]{ricker2014}, fewer than a dozen young planets have been detected and confirmed \citep[e.g.][]{benatti2019,newton2019,rizzuto2020, carleo2021a}. Therefore, it is crucial that attempts be made to fully characterize these planets particularly when they orbit bright, nearby host stars.

\citet{david2019, david2019a} reported the detection of three transiting planets and one candidate planet orbiting the young ($\approx20\,{\rm Myr}$), bright ($V=10\,{\rm mag}$) T-Tauri star, V1298 Tau, which was observed during \emph{K2} Campaign 4 in 2016 \citep{howell2014}. The planets have short orbital periods ($\approx8-60\,{\rm d}$) and radii between that of Neptune and Jupiter, implying that they currently host substantial H/He-dominated atmospheres that could be substantially stripped as they evolve \citep{poppenhaeger2020}. Additional transits were observed in 2021 with \emph{TESS} \citep{feinstein2022} including a second transit of planet e, which was only previously observed once by \emph{K2}. Recent mass constraints inferred from radial velocity (RV) measurements were published by \citet{suarezmascareno2021}; these measurements estimated that planets b and e exhibit masses of $\approx0.6\,M_{\rm Jup}$ and $1.2\,M_{\rm Jup}$, respectively, and that planets c and d have masses $\lesssim0.3\,M_{\rm Jup}$. The reported mass of planet e is particularly surprising since it suggests that Jupiter-mass planets may contract much more rapidly than is predicted by planetary evolution models \citep{fortney2007,baraffe2008}. However, the $40.2\,{\rm d}$ period of planet e inferred by this study is incompatible with the timing of the \emph{K2} and \emph{TESS} transits \citep{feinstein2022}, which suggests that the mass constraint needs to be revised.

Young systems such as V1298~Tau are particularly challenging targets---both for transit and RV studies---due to the high degree of stellar activity exhibited by their host stars \citep{ibanezbustos2019,gilbert2022b}. Nonetheless, previous RV studies have demonstrated the feasibility of detecting and characterizing these planetary RV signatures with the aid of Gaussian Processes (GPs) \citep{cloutier2019a,plavchan2020,cale2021,klein2021}. In this work, we applied this technique to a joint transit-RV modelling analysis using \emph{K2} and \emph{TESS} photometry, published RV measurements, and new RV measurements obtained using the MAROON-X spectrograph \citep{seifahrt2018,seifahrt2020,seifahrt2022}. The goal of this study is to better constrain the planetary masses of V1298~Tau's four transiting planets. In Sect. \ref{sect:obs}, we describe the photometric and RV measurements that were taken and included in our analysis. In Sections \ref{sect:analysis} and \ref{sect:results} we present the methods with which the analysis was carried out and the resulting planetary property constraints (mass, radius, etc.). In Section \ref{sect:disc}, we discuss the results and their potential implications for theories of planetary formation, evolution, and atmospheric mass loss.

\section{Observations}\label{sect:obs}

\subsection{Photometry}\label{sect:phot}

Multiple transits of V1298~Tau b, c, and d and a single transit of planet e were previously detected during \emph{K2} Campaign 4 \citep{david2019a, david2019}. The data set consists of 3,397 data points obtained over a $71$-day interval from 2015 February 8 to 2015 April 20 with exposure times of $\approx29\,{\rm min}$. We include these measurements in our analysis via the \texttt{EVEREST 2.0} \emph{K2} light curve \citep{luger2018} that was subsequently cleaned and published by \citet{suarezmascareno2021}. 

\citet{feinstein2022} reported the detection of the transits of V1298~Tau~b, c, d, and e using the publicly available \emph{TESS} Sectors 43 and 44 data sets \citep{ricker2014}. The measurements span an $\approx50$-day time period from 2021 September 16 to 2021 November 5  with cadences of $20\,{\rm sec}$ and $2\,{\rm min}$. We used the $2\,{\rm min}$ cadence PDCSAP\_FLUX light curve \citep{jenkins2016} available on the MAST archive\footnote{\url{https://mast.stsci.edu/portal/Mashup/Clients/Mast/Portal.html}}, which, after removing all data points with NaN values and with quality flags $\geq10$, consists of 31,341 measurements. Each of the four segments were then roughly detrended individually using a linear fit. The light curve was then binned into $10\,{\rm min}$ bins yielding a total of 6,279 data points. We searched the binned light curve for significant flares by-eye and ultimately masked out 5 points associated with a single flare event occurring at ${\rm BJD}=2459492.325$.

In addition to the \emph{K2} and \emph{TESS} light curves we also used the $V$-band light curve obtained with the Las Cumbres Observatory (LCOGT) network and published by \citet{suarezmascareno2021}. The light curves consist of 251 measurements obtained with a cadence of $8\,{\rm hrs}$ from 2019 October 26 to 2020 March 22. The photometric precision is reportedly $\sim10\,{\rm ppt}$ and is used to provide additional constraints on the stellar activity.

\subsection{Radial Velocities}\label{sect:rv}

The radial velocity measurements included in this work were obtained using 5 instruments. A total of 261 measurements published by \citet{suarezmascareno2021} were obtained from 2019 March 1 to 2020 March 29 using HARPS-N (135 measurements with a median uncertainty of $\tilde{\sigma}_{\rm RV}=8.9\,{\rm m/s}$), CARMENES (33 measurements; $\tilde{\sigma}_{\rm RV}=14.8\,{\rm m/s}$), SES (57 measurements; $\tilde{\sigma}_{\rm RV}=117\,{\rm m/s}$), and HERMES (36 measurements; $\tilde{\sigma}_{\rm RV}=50.2\,{\rm m/s}$) \citep[for further details see][]{suarezmascareno2021}. We also include 48 new spectroscopic measurements obtained from 2021 August 12 to 2021 November 23 using the MAROON-X spectrograph \citep{seifahrt2018,seifahrt2020,seifahrt2022} installed at the Gemini North telescope. Two sets of radial velocity measurements were derived from the blue and red arms of the instrument using SERVAL \citep{zechmeister2018}. The analysis yielded median RV precisions of $5.9\,{\rm m/s}$ and $10.3\,{\rm m/s}$ for the blue and red arms, respectively.

\section{Analysis}\label{sect:analysis}

A joint modelling analysis of the photometric and RV measurements (described in Sect. \ref{sect:obs}) was carried out using tools built into the \texttt{exoplanet} Python package \citep{foreman-mackey2021}. The adopted two-component models consist of (1) a GP to account for the stellar activity and instrumental noise along with (2) models for the planet-induced signatures (transits or stellar reflex RV variations).

\subsection{Photometric Modelling}\label{sect:LC}

The stellar activity and instrumental noise was modeled using a GP implemented with \texttt{celerite2} \citep{foreman-mackey2017,Foreman_Mackey_2018}. We adopted a kernel consisting of two stochastically driven damped simple harmonic oscillator (SHO) terms centered on the stellar rotation period ($P_{\rm rot}$) and it's first harmonic. This kernel is characterized by a power spectral density \citep[Eqn. 20 of ][]{foreman-mackey2017} given by
\begin{equation}\label{eqn:GP_SHO}
\begin{aligned}
    S(\omega)=&\sqrt{\frac{2}{\pi}}\sum_{n=1}^2\frac{S_n\omega_n^4}{(\omega^2-\omega_n^2)^2+\omega_n^2\omega^2/Q^2}\\
\end{aligned}
\end{equation}
where
\begin{equation}
S_1=\frac{A^2}{\omega_1Q},\;\;\;S_2=f_{\rm mix}\frac{A^2}{\omega_2Q},\;\;{\rm and}\;\;\omega_n=\frac{2\pi n}{P_{\rm rot}}.
\end{equation}
Here, $\omega_n$ corresponds to the undamped angular frequency of the oscillations; $S_1$ and $S_2$ determine the amplitude of the oscillations at $P_{\rm rot}$ and it's first harmonic where the latter is set with respect to the former using the $f_{\rm mix}$ parameter. The quality factor $Q$ describes how quickly the oscillations will die off where $0<Q<1/2$ leads to overdamped oscillations and a broad power spectral density while $Q>1/2$ leads to underdamped oscillations and a sharper power spectral density \citep[see Fig. 1 of ][]{foreman-mackey2017}. Following \citet{david2019a}, we force the kernel to be underdamped by reparameterizing $Q$ as
\begin{equation}
Q=\frac{1}{2}+Q_0
\end{equation}
and sampling $Q_0$ in log space. The diagonal elements of the covariance matrix include contributions from the individual measurement uncertainties ($\sigma_i$) and a jitter term that accounts for additional sources of white noise ($\sigma_{\rm jit}^2$), which are added in quadrature (i.e., $\sqrt{\sigma_i^2+\sigma_{\rm jit}^2}$).

Unique $A$, $f_{\rm mix}$, $P_{\rm rot}$, $Q_0$, and $\sigma_{\rm jit}$ hyperparameters were assigned to each light curve. We fixed the $\sigma_{\rm jit}$ parameter assigned to the \emph{K2} photometry to the in-transit white noise level estimated by \citet{david2019a} of $360\,{\rm ppm}$. For the \emph{TESS} photometry, we use a fixed jitter based on a conservative estimate of the in-transit jitter of $850\,{\rm ppm}$. This was estimated by first fitting the \emph{K2} and \emph{TESS} light curves individually while including $\sigma_{\rm jit}$ as a free parameter. The maximum a posteriori (MAP) solution yielded $\sigma_{\rm jit}^{TESS}/\sigma_{\rm jit}^{K2}\approx2.36$, which was then used to estimate the in-transit \emph{TESS} noise level of $\sigma_{\rm jit}^{TESS}=850\,{\rm ppm}$. No transits are detectable in the $LCOGT$ photometry due to the lower precision and longer cadence; therefore, $\sigma_{\rm jit}^{LCOGT}$ was set as a free parameter.

The planetary transit component was generated using the \texttt{starry} analytic light curve model \citep{luger2019}. The model is parameterized by the stellar mass ($M_\star$) and radius ($R_\star$) along with each planet's orbital period ($P$), mid-transit time ($T_0$), planet-star radii ratio ($R_{\rm p}/R_\star$), impact parameter ($b$), eccentricity ($e$), and argument of periastron ($\omega$). Unique sets of limb darkening constants ($u_1$ and $u_2$) are used for the \emph{K2} and \emph{TESS} light curves \citep[sampled using the $q_1$ and $q_2$ parameterization recommended by][]{kipping2013}. For the $e\geq0$ model, $e$ and $\omega$ are reparameterized by sampling in $\sqrt{e}\cos\omega$ and $\sqrt{e}\sin\omega$ where $\omega$ corresponds to the host star; we then applied a prior on the eccentricity based on the empirical multi-planet $e$ distribution published by \citet{vaneylen2019}.

In addition to each light curve's set of GP hyperparameters, we also include zero-point offset terms ($\langle f\rangle$). \emph{TESS} has a significantly larger pixel size relative to \emph{Kepler} ($\approx21\,{\rm arcsec/pxl}$ compared with $\approx4\,{\rm arcsec/pxl}$), which can potentially introduce contamination from background sources that may alter the transit depths measured between the two instruments. In order to account for this, we initially included a flux-dilution term that scales the planetary component flux associated with the \emph{K2} light curve; however, no evidence of dilution was found based on this factor being $\sim1$ so the parameter was removed from the subsequent fits.

\subsection{Radial Velocity Modelling}\label{sect:RV}

Modelling of the five RV data sets was carried out using the same framework that was used for the photometric modelling: the stellar activity and instrumental noise was modelled using GPs in conjunction with models describing the planetary contributions to the stellar radial velocity variations. As noted by previous RV studies of young, active systems, the accuracy with which the stellar activity can be modeled with GPs can be sensitive to the choice of covariance function \citep[e.g.][]{benatti2021,suarezmascareno2021}. Based on injection-recovery tests that we carried out (see Sect. \ref{sect:injection} in the Appendix), we found that the SHO kernel (Eqn. \ref{eqn:GP_SHO}) yielded relatively poor accuracy for planets b, d, and e and significantly under-estimated the semi-amplitudes of planet d's injected signals. The highest overall accuracy was achieved by adopting the Quasi-Periodic kernel \citep[Eqn. 3.21 of][]{roberts2013}:
\begin{equation}\label{eqn:GP_QP}
\begin{aligned}
    k(\tau)=A^2\exp\left[-\frac{\sin^2(\pi\tau/P_{\rm rot})}{2\lambda_p^2}-\frac{\tau^2}{\lambda_e^2}\right]
\end{aligned}
\end{equation}
where $\tau$ is the difference in time between any two data points, $A$ is the amplitude, $\lambda_p$ is a dimensionless length scale that specifies the complexity of the periodic variations (lower $\lambda_p$ implies greater complexity), and $\lambda_e$ is the exponential decay timescale. Individual measurement uncertainties and jitter terms were included as contributions to the covariance matrix diagonal elements using the same approach used for the photometric GP activity model. Each RV data set was assigned unique $A$ and $\sigma_{\rm jit}$ terms. Two sets of $P_{\rm rot}$, $\lambda_p$, and $\lambda_e$ terms were used: one set was assigned to the RV measurements published by \citet{suarezmascareno2021} and the second set were assigned to the MAROON-X red and blue arm measurements, which were obtained $\approx2\,{\rm yrs}$ after the HARPS and CARMENES measurements.

The stellar reflex RV variations induced by each planet's Keplerian orbit was modelled using the \texttt{RadVel} Python package \citep{fulton2018a} incorporated into \texttt{exoplanet}. The models are parameterized using the systemic or mean center-of-mass velocity ($\gamma_0$), the planet mass $M_{\rm p}$ (converted into an RV semi-amplitude based on the specified $M_\star$, $P$, and $e$), $P$, $T_0$ (converted to the time of periastron), $e$, and $\omega$.

\subsection{NUTS HMC Sampling}\label{sect:mcmc}

Posterior distributions for the various model parameters were derived using the Hamiltonian Markov chain (HMC) based No-U-Turn Sampler (NUTS) algorithm \citep{hoffman2014}. This was carried out by initializing two chains and adapting the step sizes for a target acceptance rate of $0.95$ using 2000 tuning steps (the acceptance rate was increased to $0.97$ for the eccentric orbit modelling in order to avoid divergences). After discarding the tuning steps, $10000$ draws were made yielding a total of $20000$ samples combined from the two chains. Convergence was tested using the $\hat{R}$ statistic \citep{gelman1992}, which was determined to be $<1.01$ for all cases presented in this work. Priors adopted for the fitting parameters are listed in Table \ref{tbl:pl}.

\subsection{Four Planet Model}\label{sect:model}

Multiple transits of V1298~Tau~b, c, and d have been detected in both the \emph{K2} and \emph{TESS} light curves \citep{david2019a, feinstein2022}; the transit of planet e, on the other hand, was only detected once in each of these data sets. \citet{suarezmascareno2021} report the detection of an RV signal having a period of $40.2\pm1.0\,{\rm d}$ and a semi-amplitude of $62_{-16}^{+15}\,{\rm m/s}$, which they attribute to planet e. While the $40.2\,{\rm d}$ period is consistent with the detected \emph{K2} transit of planet e, it is inconsistent with the new period lower limit placed by the \emph{TESS} transit. \citet{feinstein2022} note that the possible orbital periods for e have a lower bound of $P_{\rm e}>42.7\,{\rm d}$, which corresponds to the time between the observed \emph{TESS} transit and the last \emph{TESS} measurement. Therefore, assuming circular Keplerian orbits (i.e., ignoring any transit timing variations (TTVs)) and considering only the timing of the observed transits, there are a total of 55 \emph{discrete} solutions for planet e's period given by
\begin{equation}\label{eqn:Pe}
    P_{\rm e}\approx\Delta T_{\rm e}/n_{\rm e}\;\;\;{\rm for}\;\;\;n_{\rm e}=1,\dots,55
\end{equation}
where $\Delta T_{\rm e}$ is defined as the difference between the \emph{TESS} and \emph{K2} transit times (${\rm BJD}_{\rm e}^{TESS}-{\rm BJD}_{\rm e}^{K2}\approx6.5\,{\rm yrs}$) and $n_{\rm e}=55$ corresponds to the shortest period that remains $>42.7\,{\rm d}$.

We derived solutions for all of the $P_{\rm e}$ values defined by Eqn. \ref{eqn:Pe} using all of the available RV and photometric data sets. This was done by adopting narrow priors centered on each $P_{\rm e}$ value being considered\footnote{The impact of the chosen narrow priors in both $P$ and $T_0$ (see Table \ref{tbl:pl}) was tested by increasing the width of these priors by a factor of 10; the derived posteriors were not found to be noticeably impacted.}. For expediency, we initially assumed circular orbits, which we note causes planet e's impact parameter ($b_{\rm e}$) to increase with decreasing $n_{\rm e}$ such that $b_{\rm e}$ approaches $1$ for $n_{\rm e}\lesssim20$ ($P_{\rm e}\gtrsim119\,{\rm d}$).

\subsection{Transit Times}\label{sect:TTs}

Individual transit times associated with each of the \emph{K2} and \emph{TESS} transits identified by \citet{david2019a} and \citet{feinstein2022}, respectively, were derived using the general framework described above. The two data sets were modelled simultaneously (without the inclusion of the RVs or the $LCOGT$ light curve) using the \texttt{TTVOrbit} class of the \texttt{exoplanet} package, which introduces an additional free parameter for each transit event that specifies it's time of occurrence. We adopted normal prior distributions for each transit time with mean values calculated using the published \emph{K2} ephemerides \citep{david2019a} and \emph{TESS} ephemerides \citep{feinstein2022} and with a standard deviation of $0.05\,{\rm d}$; we tested whether the derived posteriors are sensitive to the adopted prior width by increasing this width by a factor of 10, which did not have a noticeable impact. For planet e, transit indices were specified in the \texttt{TTVOrbit} class assuming $n_{\rm e}=51$ ($P_{\rm e}\approx44\,{\rm d}$ in Eqn. \ref{eqn:Pe}), which corresponds to the most probable $n_{\rm e}$ derived later in the analysis (Sect. \ref{sect:Pe_constraints} below). The transit times were initially estimated assuming $n_{\rm e}=54$, which yielded similar results.

\section{Results}\label{sect:results}

\subsection{Transit Times}\label{sect:results_TTs}

\begin{figure}
	\centering
	\includegraphics[width=1\columnwidth]{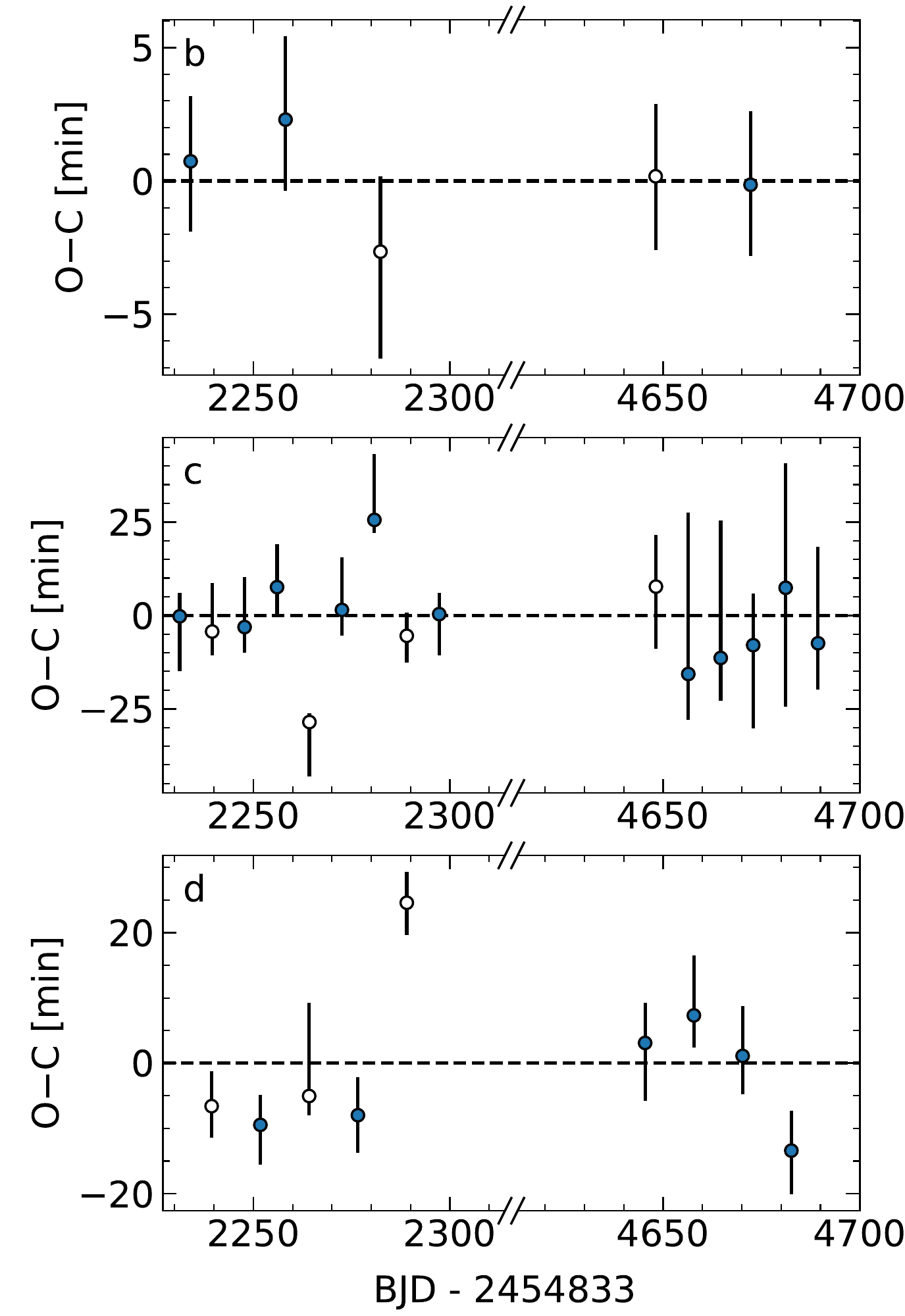}\vspace{-0.1cm}
	\caption{O$-$C values for the derived transit times measured with respect to each planet's $\langle P\rangle$; the break in the $x$-axis separates the \emph{K2} measurements (${\rm BJD}-2454833<2300$) from the \emph{TESS} measurements (${\rm BJD}-2454833>4640$). White points are those measurements that may be biased due to overlapping transits or partial event coverage.}
	\label{fig:oc}
\end{figure}

In Table \ref{tbl:tt} of the Appendix, we list the transit times and observed-minus-calculated (O$-$C) values determined from the derived posterior distributions. Reported values correspond to each distribution's median value and the uncertainties correspond to $15.9$ and $84.1$ percentiles. Average uncertainties in the transit times for planets b, c, d, and e range from $\approx0.003-0.012\,{\rm d}$; O$-$C uncertainties of planets b, c, and d are approximately $3$, $16$, and $8\,{\rm min}$, respectively.

In Fig. \ref{fig:oc}, we show the O$-$C values associated with the derived transit times for planets b, c, and d. Some of the transit times and their estimated uncertainties are impacted by biases that can be attributed to instances of coincident transit events (e.g., planet c's fifth transit and d's third transit in the \emph{K2} data) or partial event coverage (e.g., planet b's third transit). In total, $9$ of the $31$ transit times derived from the \emph{K2} and \emph{TESS} photometry may be affected by such biases (2 for planet b, 4 for planet c, and 3 for planet d). No clear evidence of TTVs are obtained from our analysis (regardless of whether or not the potentially biased transit times are considered), which is consistent with the findings of \citet{david2019a} and \citet{feinstein2022}. As a result, the joint transit-RV modelling analysis presented below, which was conducted using all of the publicly available data sets and the new MAROON-X data, assumed Keplerian orbits and used Gaussian priors for the orbital periods centered on the mean periods derived from this transit timing analysis. We note, however, that additional transit observations do exhibit significant TTVs (J. Livingston et al. 2022, in preparation); the fact that we do not find evidence of TTVs in the \emph{K2} or \emph{TESS} light curves can be attributed to the $\approx4.5\,{\rm yr}$ super-period describing the TTVs for planets c and d as predicted by \citet{david2019a} and the fact that the two data sets were likely obtained during low TTV amplitude phases of this super-period. The ephemerides reported in this work should therefore be used with caution in regards to future transit timing predictions.

\subsection{Orbital Period of Planet e}\label{sect:Pe_constraints}

\begin{figure}
	\centering
	\includegraphics[width=0.9\columnwidth]{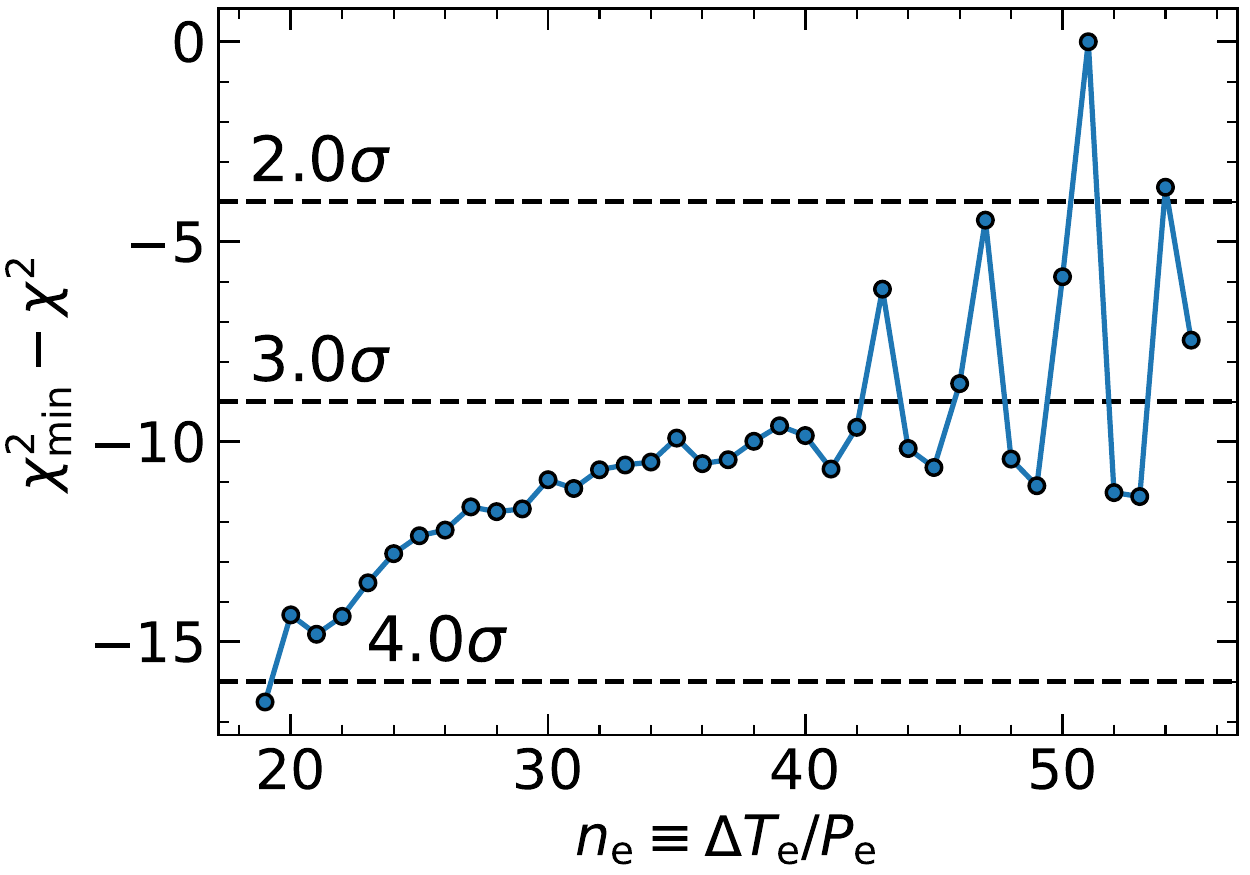}\vspace{-0.1cm}
	\caption{$\Delta\chi^2\equiv\chi^2_{\rm min}-\chi^2$ values associated with each of the $n_{\rm e}$ solutions (Eqn. \ref{eqn:Pe}) found to be consistent with the most probable solution ($n_{\rm e}=51$, $P_{\rm e}\approx46.8\,{\rm d}$) at an $\approx4\sigma$ level; lower $n_{\rm e}$ values, which decrease monotonically towards $\chi^2_{\rm min}-\chi^2<-90$ at $n_{\rm e}=1$, are not shown for visual clarity. The $2\sigma$, $3\sigma$, and $4\sigma$ confidence intervals are indicated by the horizontal dashed lines. Under the assumption of circular orbits, all longer periods with $P_{\rm e}>55.4\,{\rm d}$ are rejected at a $3\sigma$ level in favor of the $n_{\rm e}=51$ solution while seven shorter period solutions with $43.3<P_{\rm e}/{\rm d}<55.4$ ($n_{\rm e}=43$, 46, 47, 50, 51, 54, and 55) are consistent within $3\sigma$.}
	\label{fig:ne_BIC}
\end{figure}

The constraints on $P_{\rm e}$ given by Eqn. \ref{eqn:Pe} are based only on the timing of the two observed transits within the \emph{K2} and \emph{TESS} light curves. In order to determine whether the joint light curve and RV analysis provides additional constraints on $P_{\rm e}$ (i.e., on $n_{\rm e}\equiv \Delta T_{\rm e}/P_{\rm e}$), we compared the $\chi^2$ values associated with each of the $36$ $P_{\rm e}$ solutions that we considered. Each $\chi^2$ value was calculated using the median of the log-likelihood distributions obtained from the sampling analysis. We found that the solution that yielded the lowest $\chi^2$ is defined by $n_{\rm e}=51$ ($P_{\rm e}=46.768131\pm0.000076\,{\rm d}$), which we adopt as the most probable value. In Fig. \ref{fig:ne_BIC}, we show the $\Delta\chi^2\equiv\chi^2_{\rm min}-\chi^2$ values for the tested $n_{\rm e}$ solutions calculated with respect to the most probable solution. The seven solutions yielding the lowest $\chi^2$ values are defined by $n_{\rm e}=43$, 46, 47, 50, 51, 54, and 55 and are consistent within $3\sigma$. For longer periods ($P_{\rm e}>55.4\,{\rm d}$), $\Delta\chi^2$ decreases approximately monotonically with decreasing $n_{\rm e}$ (increasing $P_{\rm e}$) towards $\Delta\chi^2<-90$ at $n_{\rm e}=1$. Below we report the results of the adopted $n_{\rm e}=51$ solution and how the associated mass constraints compare with those of the other six $P_{\rm e}$ solutions that cannot be ruled out from the analysis presented in this work at a $3\sigma$ level.

All of the derived parameters for the $n_{\rm e}=51$ solution are listed in Table \ref{tbl:pl}. In Figures \ref{fig:k2} to \ref{fig:tess_zoom}, we show the median solution fits (i.e., the solution calculated using the median value of each posterior that are listed in Table \ref{tbl:pl}) to the \emph{K2} and \emph{TESS} light curves obtained for the adopted $n_{\rm e}=51$ solution. The associated fits to the RV measurements are shown in Fig. \ref{fig:rv} and the phased RVs showing individual planetary contributions to the measurements are shown in Fig. \ref{fig:rv_phase}.

\subsection{Planetary Masses}\label{sect:mass_constraints}

\begin{figure}
	\centering
	\includegraphics[width=0.42\textwidth]{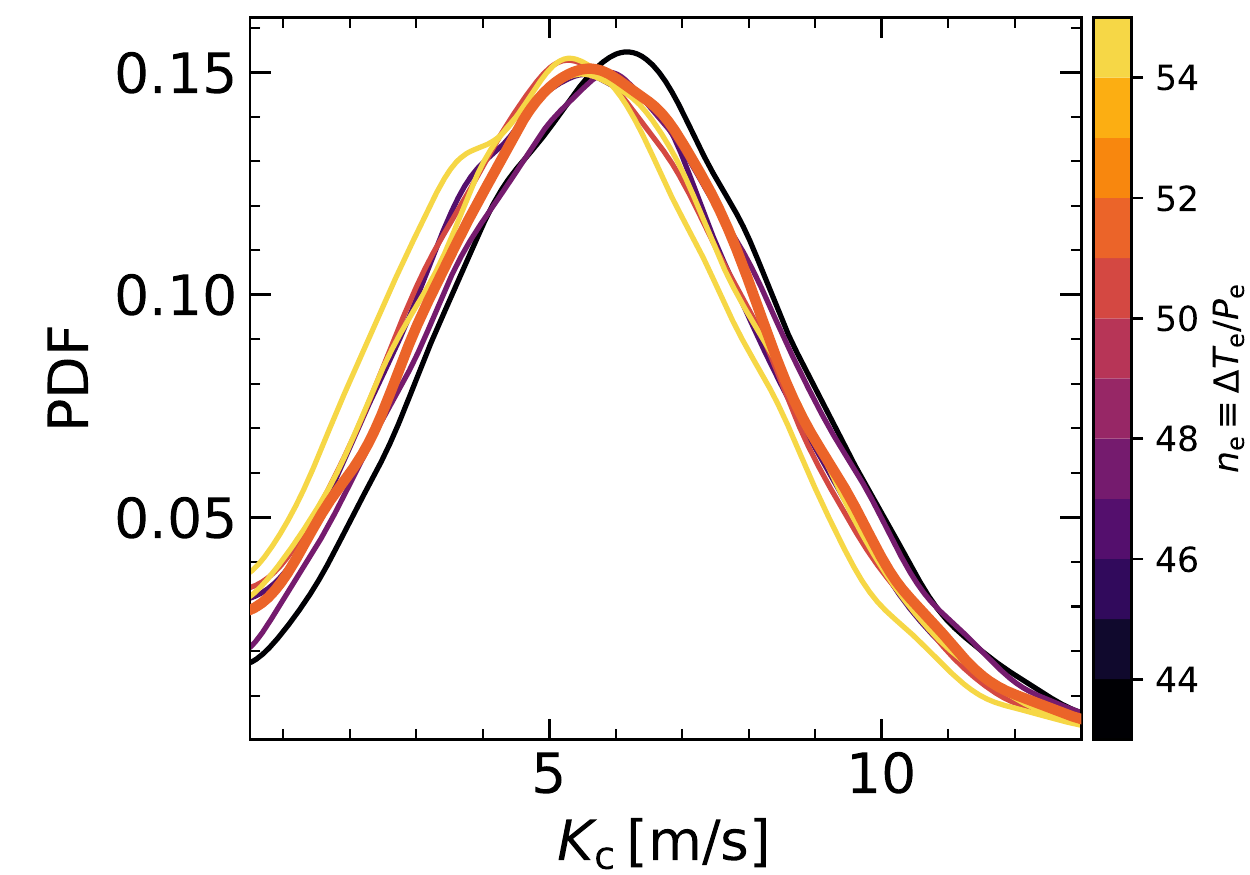}
	\includegraphics[width=0.42\textwidth]{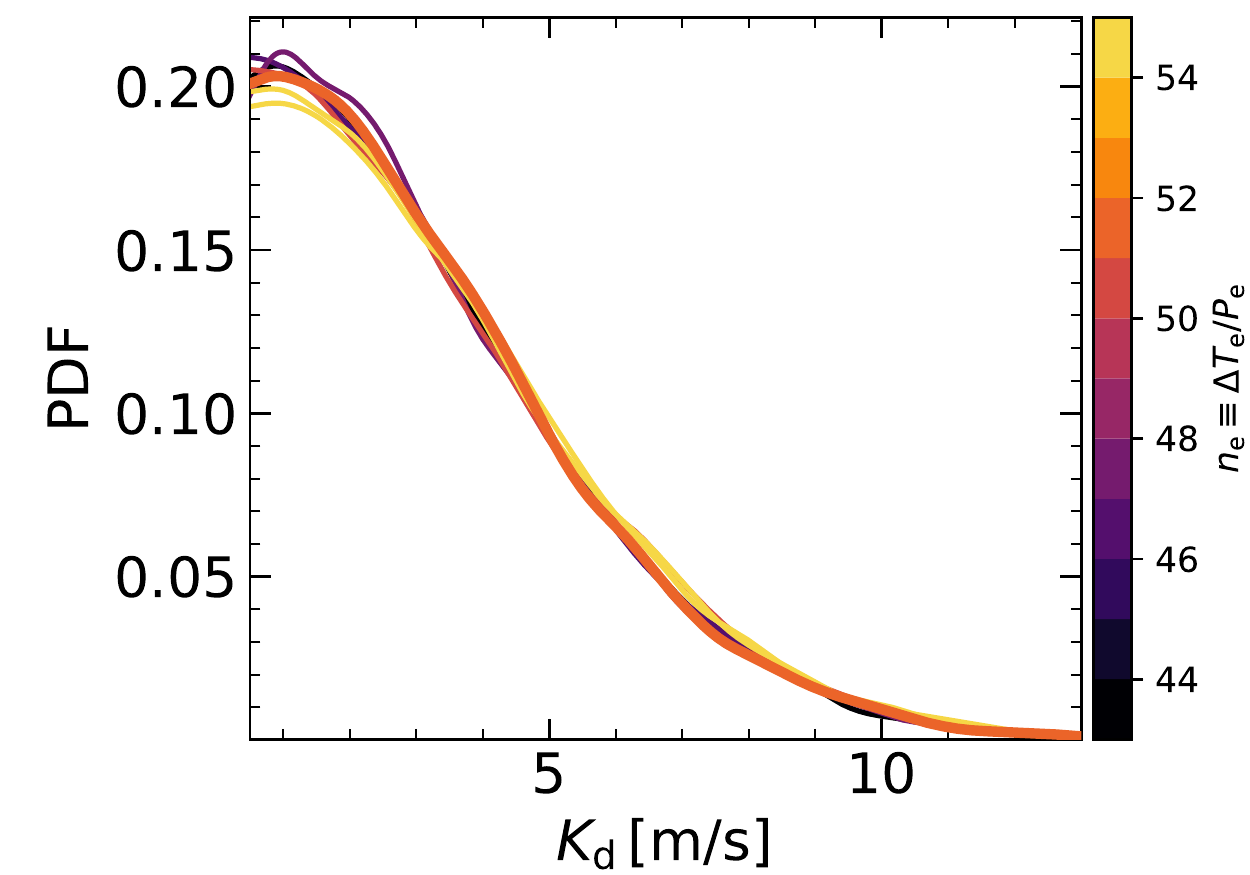}
 	\includegraphics[width=0.42\textwidth]{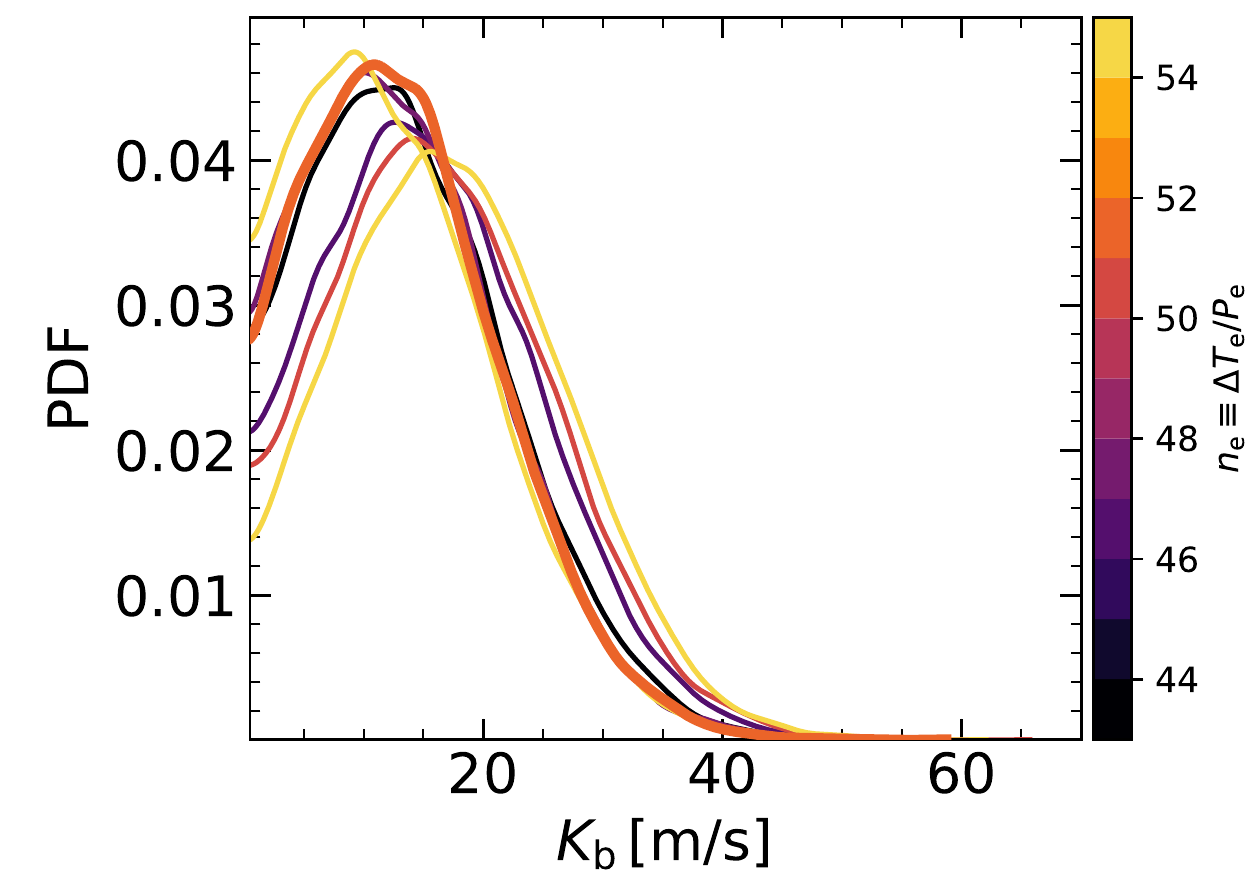}
	\includegraphics[width=0.42\textwidth]{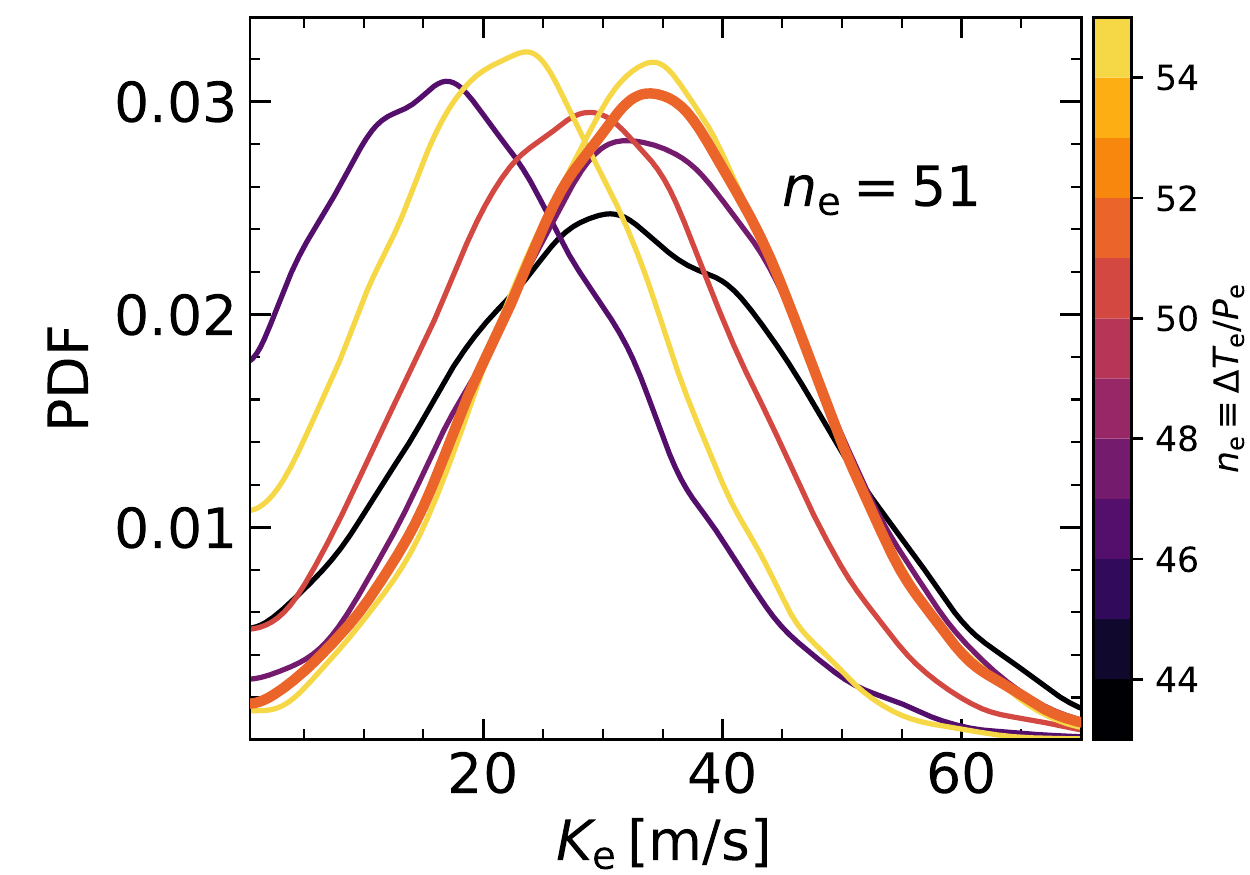}
	\caption{Marginalized posterior distributions derived for the RV semi-amplitudes of planets c, d, b, and e for the seven most probable $n_{\rm e}$ solutions---those that are consistent with the most probable $n_{\rm e}=51$ solution ($P_{\rm e}\approx46.8\,{\rm d}$) within a $3\sigma$ confidence interval. The thicker line indicates the $n_{\rm e}=51$ solution.}
	\label{fig:K_pdf}
\end{figure}

\begin{figure*}
	\centering
	\includegraphics[width=2\columnwidth]{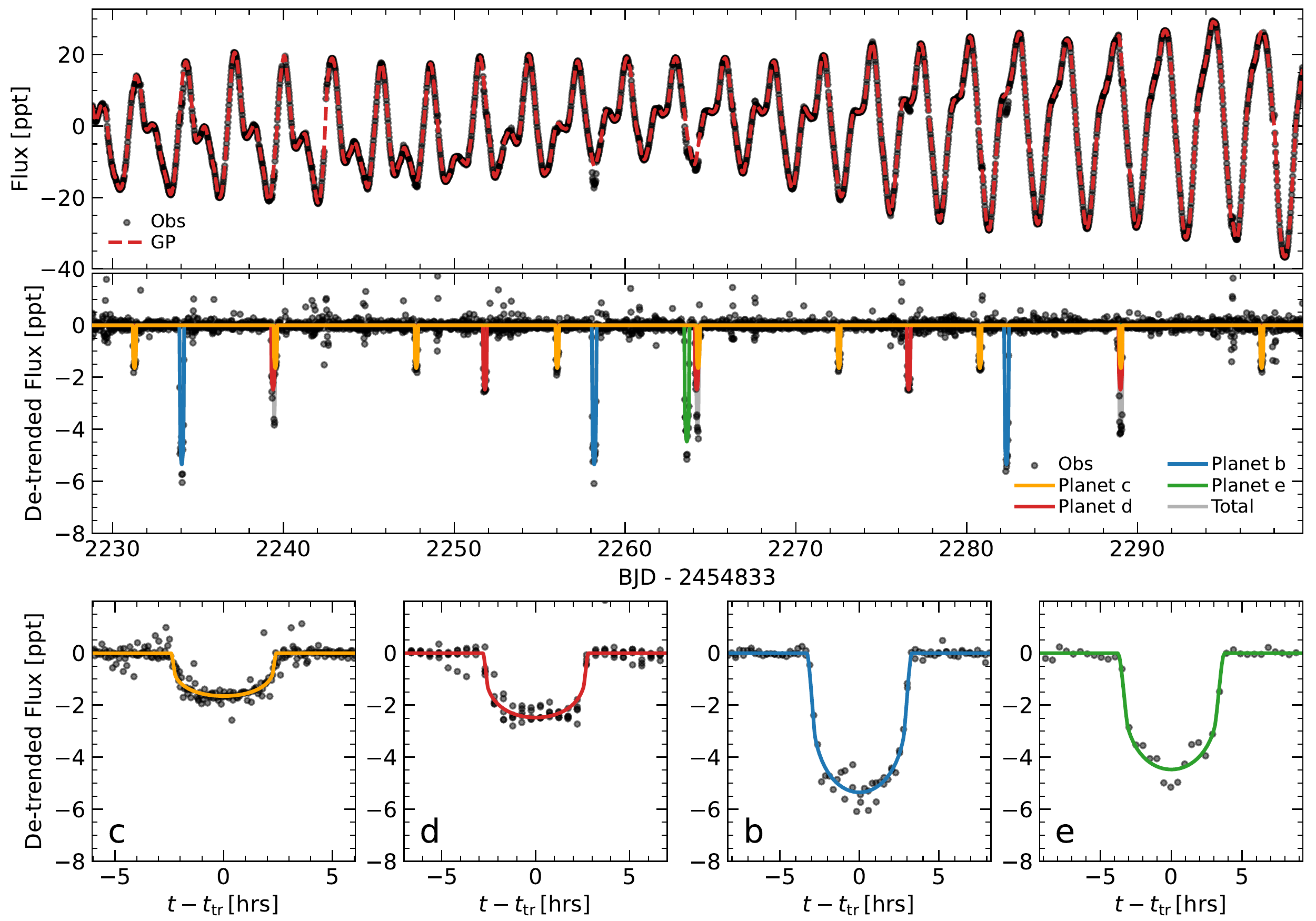}\vspace{-0.1cm}
	\caption{Best fit to the \emph{K2} light curve using the $n_{\rm e}=51$ ($P_{\rm e}\approx46.8\,{\rm d})$ solution. The top panel shows the observed flux measurements compared with the GP model. The middle panel shows the transit models compared with the de-trended measurements (i.e., with the GP model removed). The bottom 4 panels show the phased transits for planets b, c, d, and e; individual transits are shown in \ref{fig:k2_zoom}.}
	\label{fig:k2}
\end{figure*}

\begin{figure*}
	\centering
	\includegraphics[height=7.5cm]{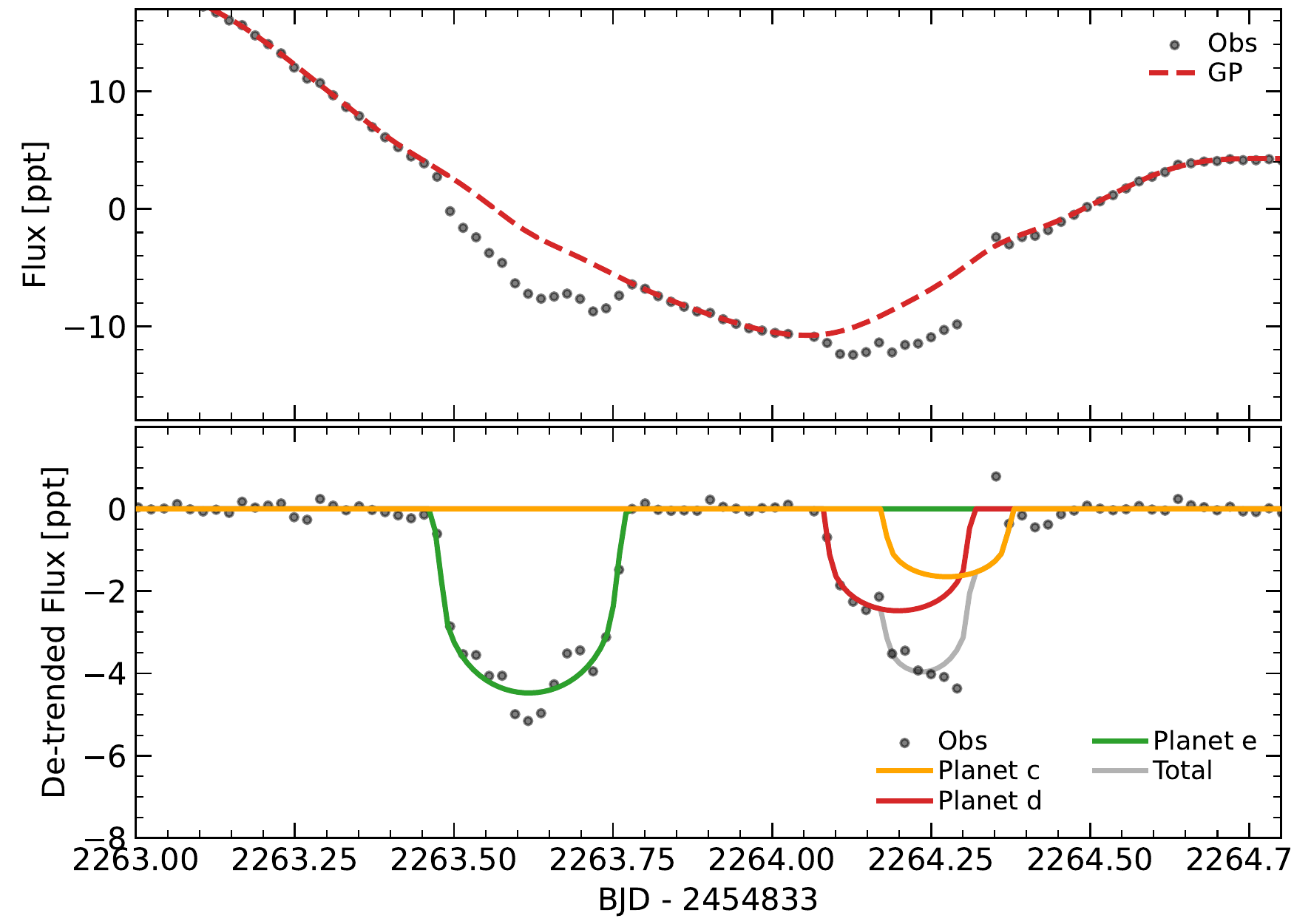}\vspace{-0.1cm}
	\includegraphics[height=7.5cm]{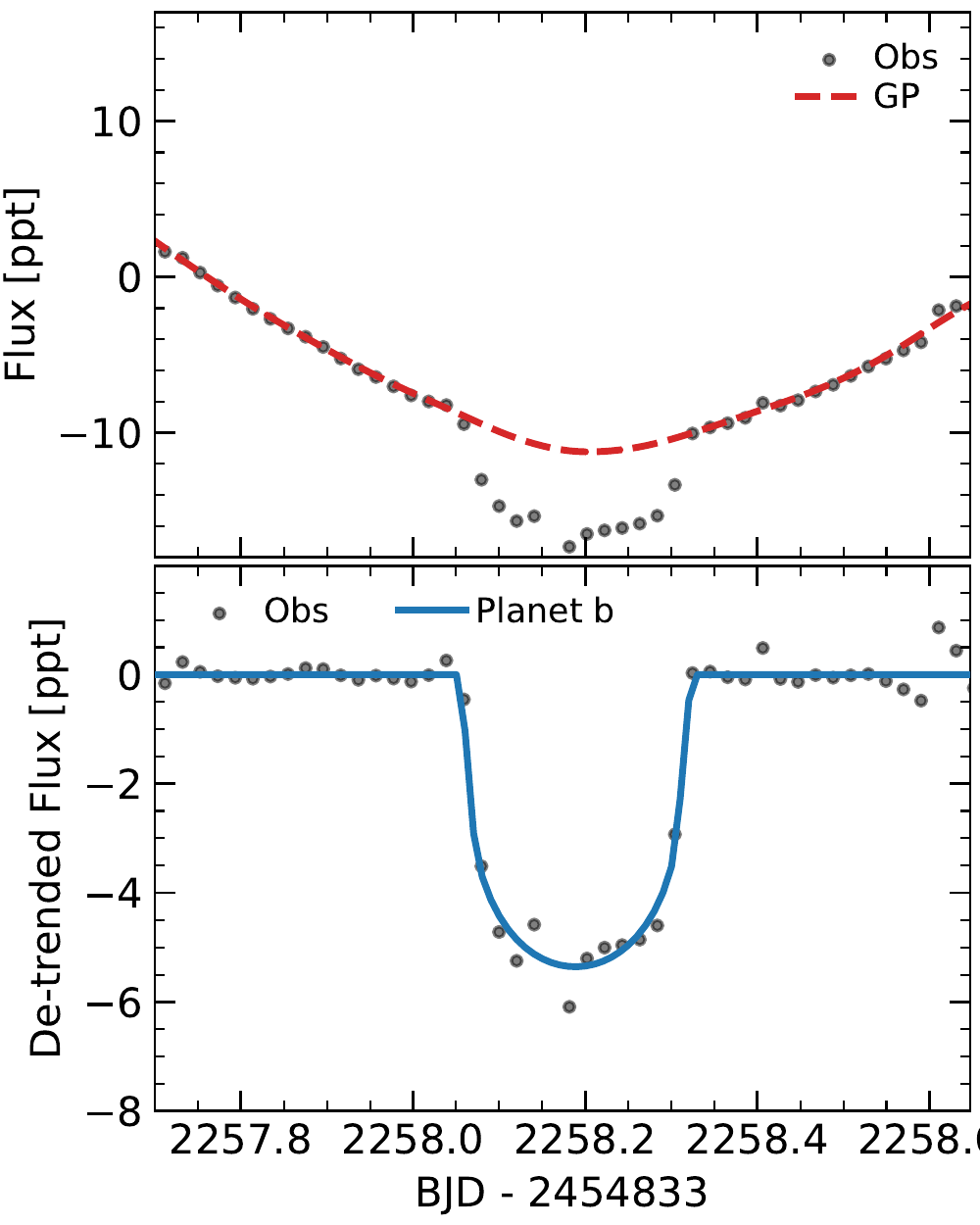}\vspace{-0.1cm}
	\caption{Selected individual transits observed in the \emph{K2} LC.}
	\label{fig:k2_zoom}
\end{figure*}

\begin{figure*}
	\centering
	\includegraphics[width=2\columnwidth]{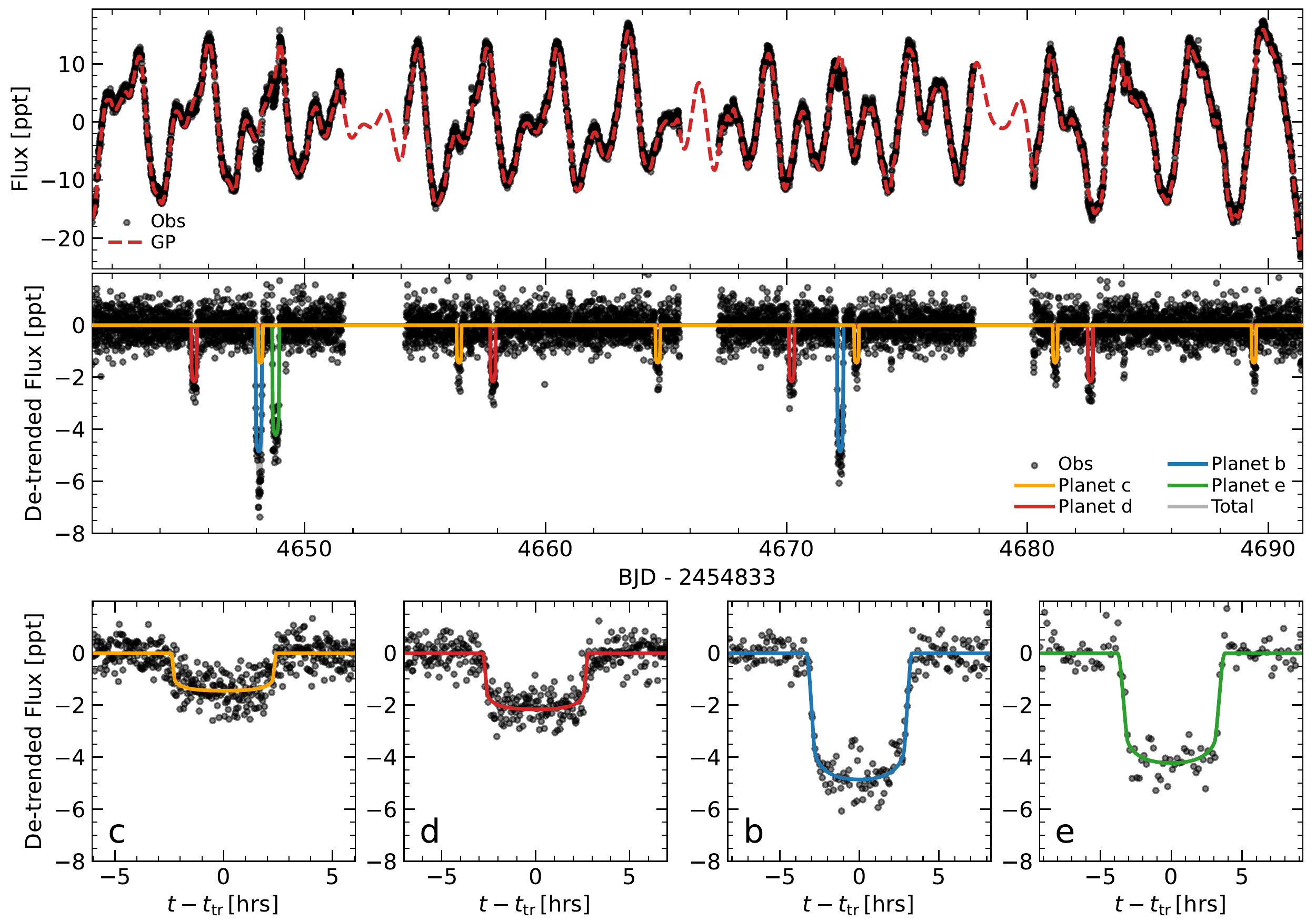}\vspace{-0.1cm}
	\caption{Same as Fig. \ref{fig:k2} but for the \emph{TESS} LC. Individual transits are shown in Fig. \ref{fig:tess_zoom}.}
	\label{fig:tess}
\end{figure*}

\begin{figure*}
	\centering
	\includegraphics[height=7.5cm]{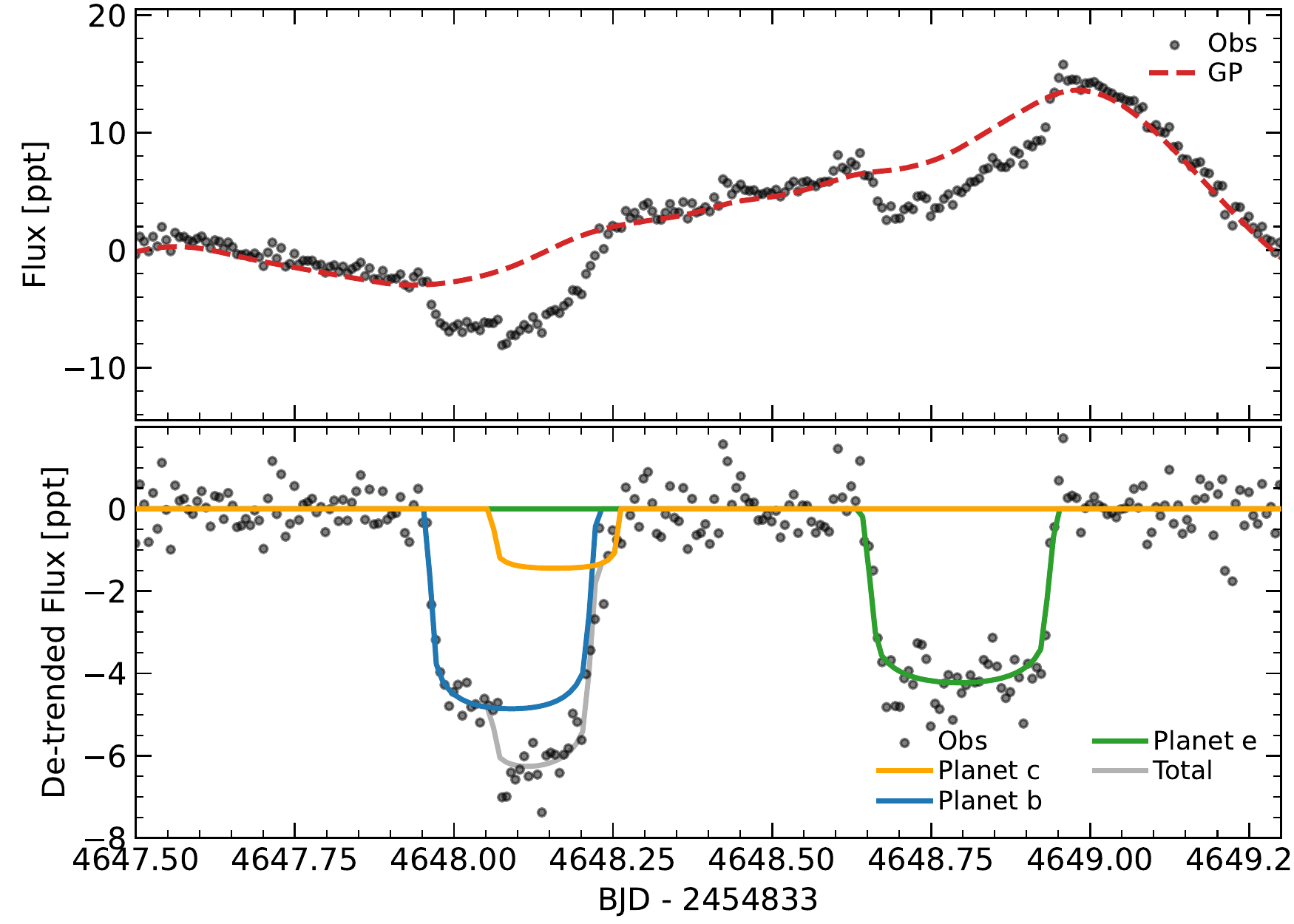}\vspace{-0.1cm}
	\includegraphics[height=7.5cm]{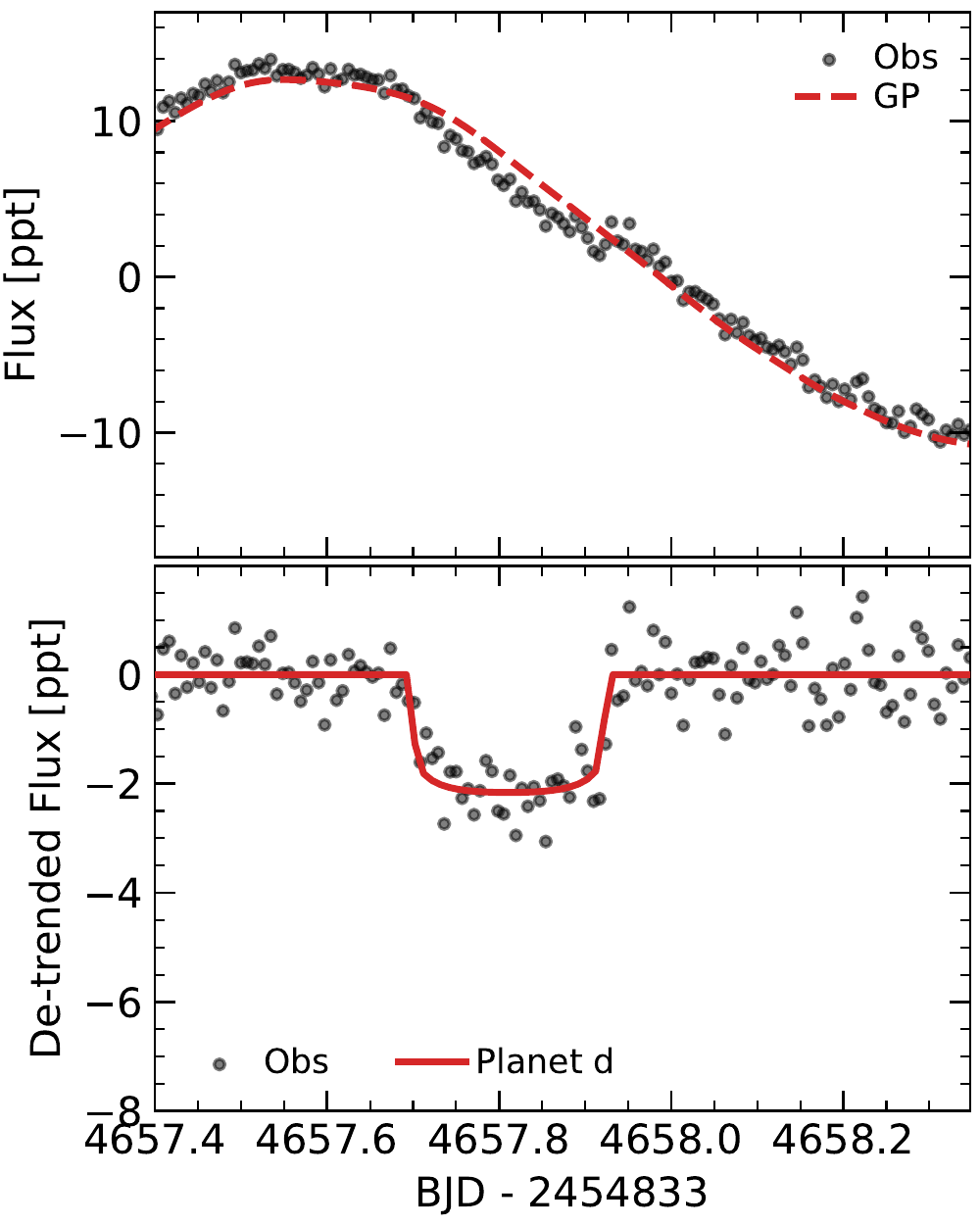}\vspace{-0.1cm}
	\caption{Selected individual transits observed in the \emph{TESS} LC.}
	\label{fig:tess_zoom}
\end{figure*}

In Fig. \ref{fig:K_pdf} we show the marginalized RV semi-amplitude posterior distributions of the four planets derived for all of the seven most probable $n_{\rm e}$ solutions (i.e., those within the $3\sigma$ confidence interval of the $n_{\rm e}=51$ solution) assuming circular orbits. The posteriors of planet e depend strongly on the adopted $n_{\rm e}$ value while those of planets b and c are moderately impacted. No clear detections of RV signatures associated with planets b and d are obtained. Weakly significant detections are obtained for planets c and e: planet c is detected with a significance of $\approx2\sigma$ (for all $n_{\rm e}$ values) and planet e has a maximum detection significance of $2.6\sigma$, which is associated with the most probable $n_{\rm e}=51$ solution. A comparable detection significance for planet e (i.e., $>2\sigma$) is also obtained for the $n_{\rm e}=43$, 47, 50, and 54 solutions, which, including the $n_{\rm e}=51$ solution, correspond to the 5 most probable solutions shown in Fig. \ref{fig:ne_BIC}.

Considering the $n_{\rm e}=51$ solution, we derive $2\sigma$ upper limits on the semi-amplitudes of planets b and d of $K_{\rm b}<32\,{\rm m/s}$ and $K_{\rm d}<9.0\,{\rm m/s}$, which correspond to upper mass limits of $M_{\rm b}<159\,M_\oplus$ and $M_{\rm d}<36\,M_\oplus$. While the injection-recovery tests that we performed (Sect. \ref{sect:injection} of the Appendix) imply that our model can accurately constrain $K_{\rm b}$, $K_{\rm c}$, and $K_{\rm e}$, they reveal a systematic bias in which $K_{\rm d}$ is underestimated by $\approx10-15\%$. Taking this bias into account implies a slightly higher upper limit on planet d's mass of $M_{\rm d}<41.4\,M_\oplus$. For planets c and e, we obtain $K_{\rm c}=5.7\pm2.6\,{\rm m/s}$ ($M_{\rm c}=19.8_{-8.9}^{+9.3}\,M_\oplus$) and $K_{\rm e}=34\pm13\,{\rm m/s}$ ($M_{\rm e}=0.66\pm0.26\,M_{\rm Jup}$). Considering the seven most probable solutions shown in Fig. \ref{fig:K_pdf}, the $n_{\rm e}$ yields the highest upper limits on the masses of planets c and e of $M_{\rm c}<39\,M_\oplus$ and $M_{\rm e}<1.34\,M_{\rm Jup}$.

\subsubsection{Constraints From Dynamical Stability}\label{sect:stability}

The dynamical stability of the adopted solution was evaluated using the Stability of Planetary Orbital Configurations Klassifier \citep[\texttt{SPOCK}; ][]{tamayo2020,tamayo2021} Python package. SPOCK is able to quickly estimate the probability that a multi-planet system with a given set of initial conditions will maintain stability over $10^9$ orbits (i.e., $\sim20\,{\rm Myr}$ for V1298 Tau). We calculated this stability probability for each of the posterior samples (i.e., using $M_\star$ along with each planet's $T_0$, $P$, inclination angle, mass, and $e$ and $\omega$ in the case of non-circular orbits) obtained from the NUTS sampling analysis. In Fig. \ref{fig:Mp_corner} of the Appendix, we show the derived $M_{\rm p}$ posteriors along with the stability probabilities calculated with \texttt{SPOCK} (black contours). The stability probability distribution is bimodal with peaks occurring at $\approx0.33$ and $\approx0.65$. The stability probabilities are most clearly anti-correlated with $M_{\rm b}$ with lower mass solutions being more stable. The blue contours show the distributions after applying rejection sampling to the stability probability distribution, which predominantly removes samples with low stability. This shifts planet b's $2\sigma$ upper mass limit down slightly to $M_{\rm b}<141\,M_\oplus$ while smaller shifts occur for the other three planets.

\subsubsection{Non-Circular Orbits}\label{sect:stability}

We carried out the same sampling analysis for the $n_{\rm e}=51$ solution presented above but allowing for non-circular orbits. In this case, we derive low eccentricities for planets b, d, and e of $e_{\rm b}<0.13$, $e_{\rm d}<0.14$, and $e_{\rm e}<0.32$. The derived masses for these planets are found to be comparable to the circular orbits case: $M_{\rm b}<149\,M_\oplus$, $M_{\rm d}<39\,M_\oplus$, and $M_{\rm e}=0.70\pm0.27\,M_{\rm Jup}$. Planet c, on the other hand, is found to have a large eccentricity of $e_{\rm c}=0.44_{-0.12}^{+0.10}$ and a notably larger mass of $39\pm11\,M_\oplus$---nearly twice that of the mass derived assuming $e=0$. We calculated the stability probabilities for the posterior samples using SPOCK and find that they have a similar distribution to that of the circular orbits case albeit with a small shift in the stability probability of $\lesssim0.05$ towards lower probabilities.

The mass and eccentricity posterior distributions along with the calculated distribution of the stability probabilities are shown in Fig. \ref{fig:Mp_corner_e} of the Appendix. Both planets c and e have bimodal eccentricity posteriors: aside from the most probable eccentricities noted above, the posteriors have peaks with lower relative probabilities at $e_{\rm c}\approx0$ and $e_{\rm e}\approx0.3$. When including only the \emph{K2} and \emph{TESS} data sets in the sampling analysis and allowing for non-circular orbits, we obtain low eccentricities for all four planets characterized by $2\sigma$ upper limits of $e_{\rm b}<0.17$, $e_{\rm c}<0.30$, $e_{\rm d}<0.12$, and $e_{\rm e}<0.25$); therefore, the high value of $e_{\rm c}$ is primarily driven by the RVs. \citet{shen2008} show that small, low signal-to-noise RV data sets may be significantly biased towards high eccentricities and high masses. Considering the semi-amplitude of planet c's RV signal ($K_{\rm c}=5.7\pm2.6\,{\rm m/s}$) and the typical measurement uncertainty of $\approx10\,{\rm m/s}$, we conclude that the lower $e_{\rm c}$ and lower $M_{\rm c}$ solution is most reliable.

\begin{figure*}
	\centering
	\includegraphics[width=0.49\textwidth]{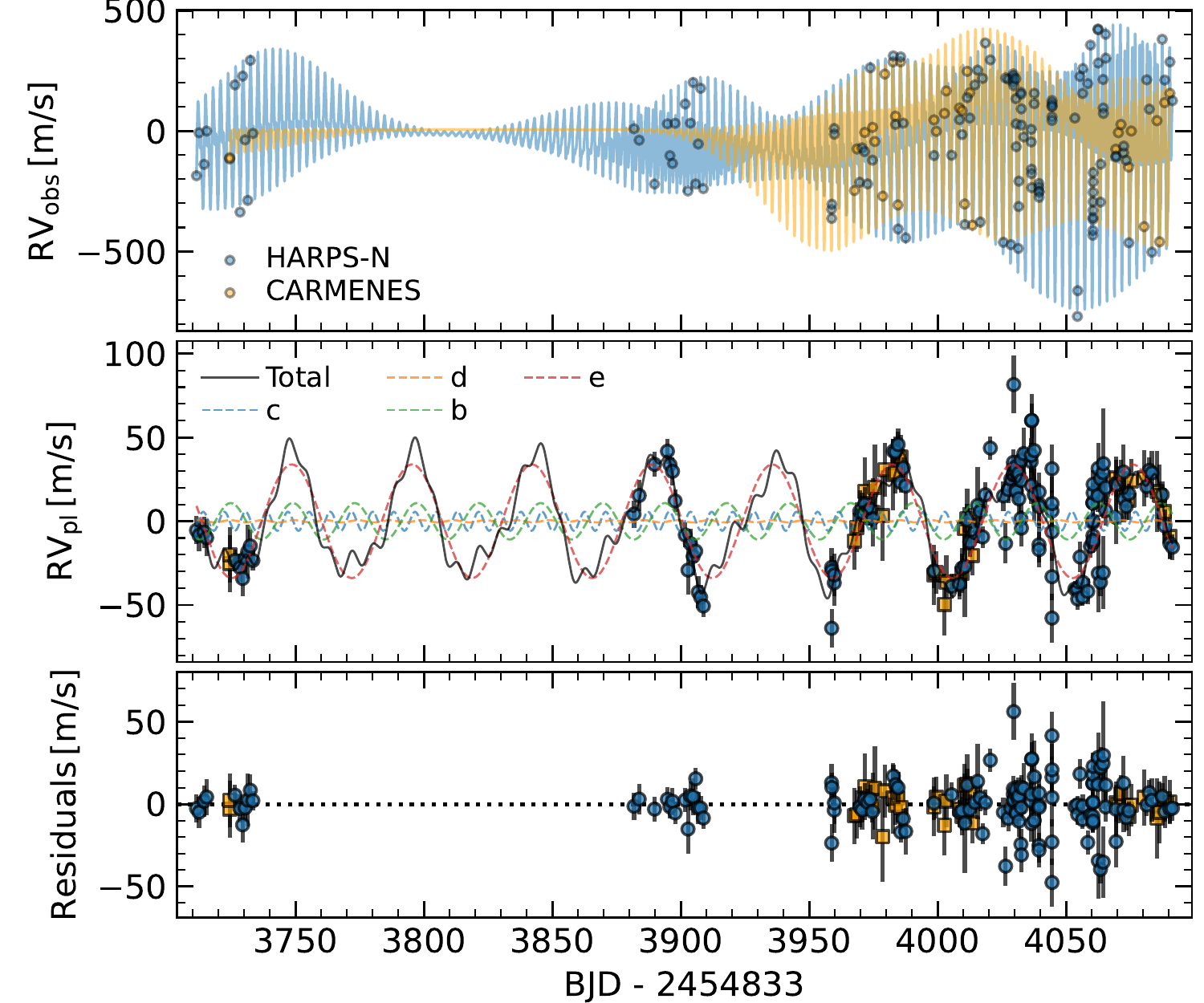}\vspace{-0.1cm}
	\includegraphics[width=0.49\textwidth]{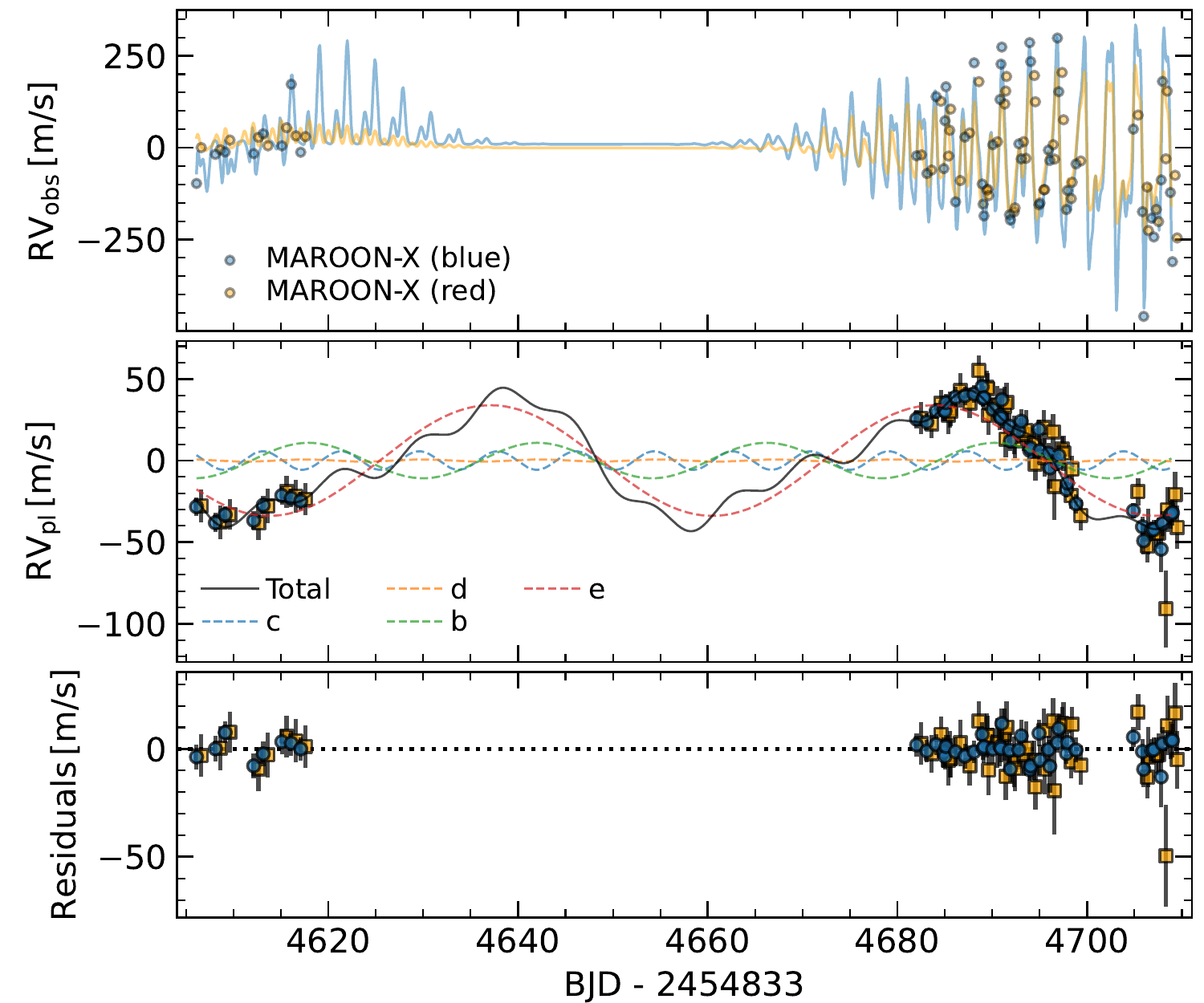}\vspace{-0.1cm}
	\caption{Best fit to the HARPS and CARMENES RV measurements (left panels) and the MAROON-X red and blue measurements (right panels) using the $n_{\rm e}=51$ ($P_{\rm e}\approx46.8\,{\rm d})$ solution (n.b., the MAROON-X measurements are plotted with a time offset for better visibility). The top panels show the GP fit to the observed RVs with the planetary component removed, the middle panels show the total planetary component with the GP fit removed, and the bottom panels show the residuals. The STELLA and HERMES measurements, which are included in our analysis, are not shown due to their comparatively large uncertainties.}
	\label{fig:rv}
\end{figure*}

\begin{figure*}
	\centering
	\includegraphics[width=0.32\textwidth]{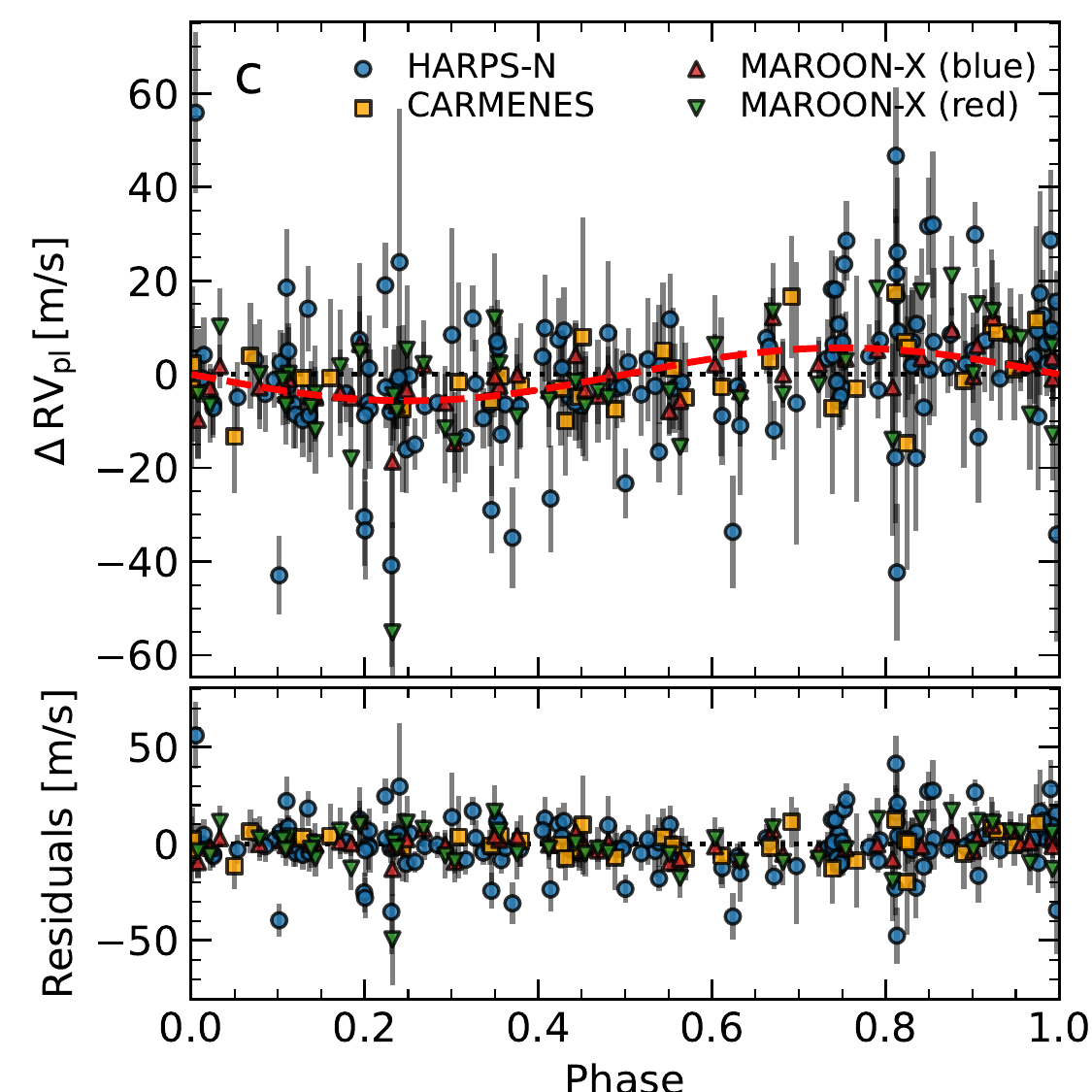}
    \includegraphics[width=0.32\textwidth]{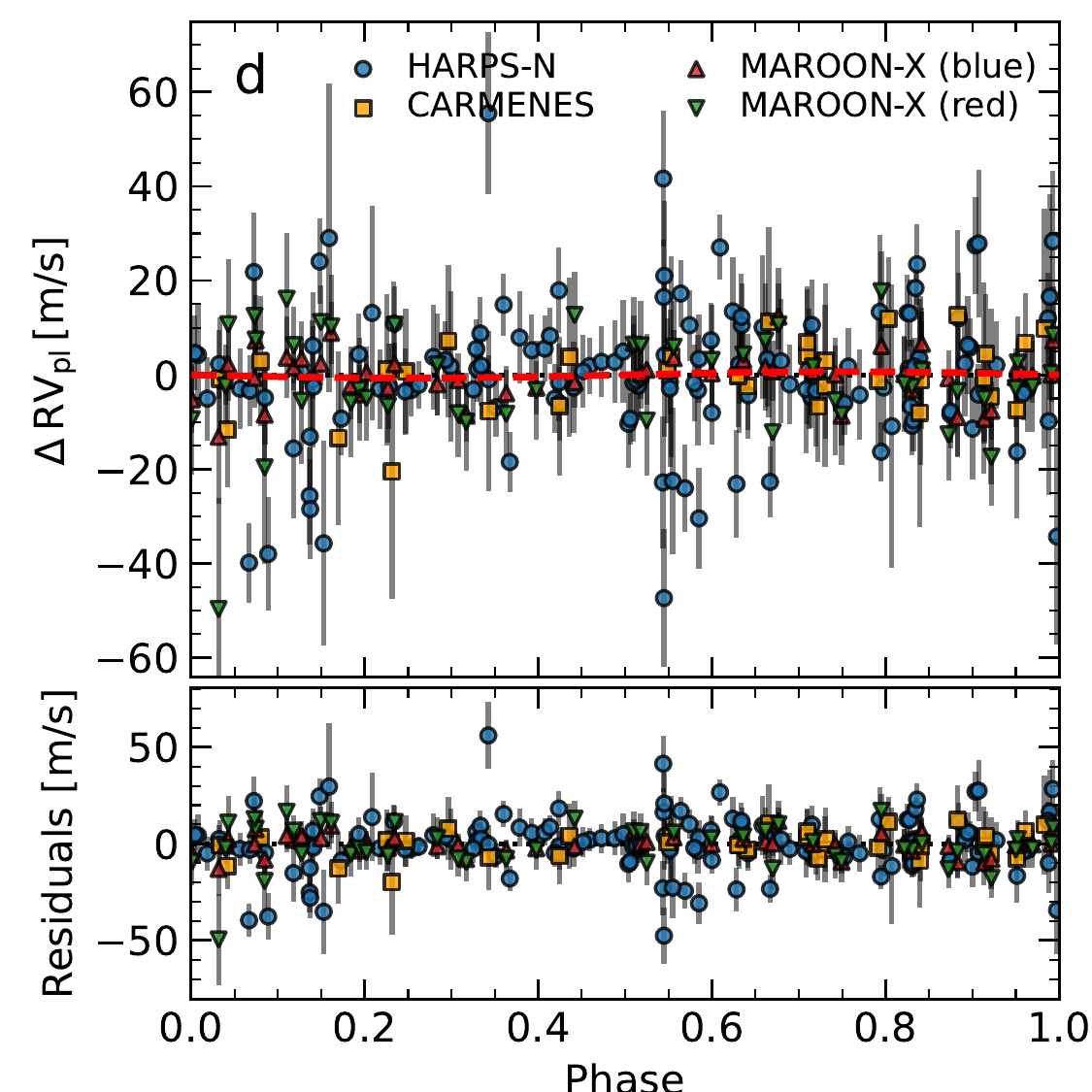}
    
	\includegraphics[width=0.32\textwidth]{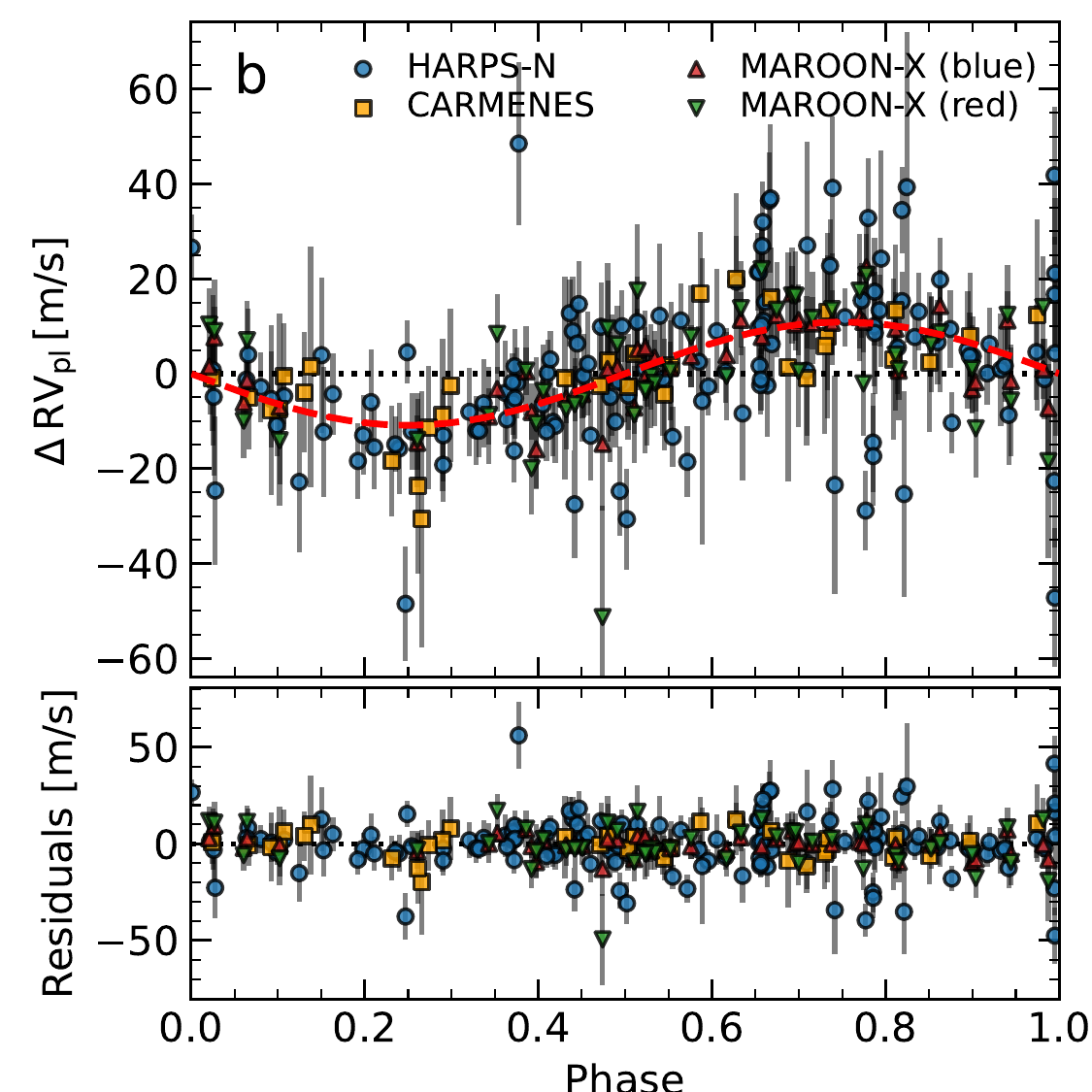}
    \includegraphics[width=0.32\textwidth]{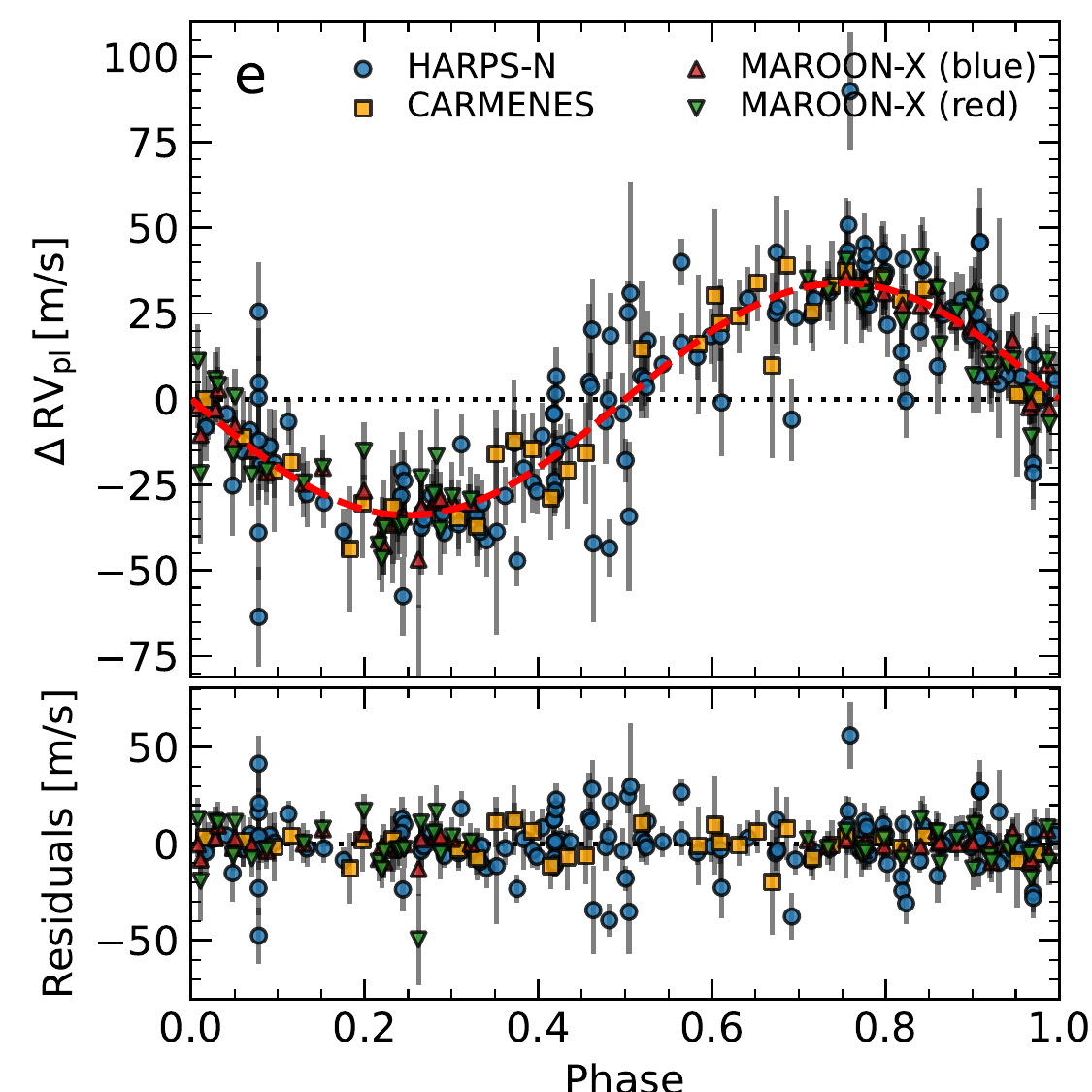}\vspace{-0.1cm}
	\caption{Individual planetary contributions to the stellar reflex RV phased by orbital period for the median solution fit (Table \ref{tbl:pl}) assuming $P_{\rm e}\approx46.8\,{\rm d}$. Filled circles correspond to the detrended observed RVs and the red dashed lines indicate the model fits. For visual clarity, the STELLA and HERMES measurements are not plotted.}
	\label{fig:rv_phase}
\end{figure*}

\section{Discussion \& Summary}\label{sect:disc}
In this study, we carried out a joint transit and RV modelling analysis of the young V1298~Tau system, which contains four transiting short-period planets with $4.9-9.6\,R_\oplus$ radii. We include the constraints imposed on planet e's orbital period by the transit observed with \emph{K2}, the transit observed with \emph{TESS}, and the RVs and ultimately obtain at least seven plausible solutions. These solutions have $43.3\,{\rm d}<P_{\rm e}<55.4\,{\rm d}$ while longer-period solutions ($43.3\,{\rm d}<P_{\rm e}<6.54\,{\rm yrs}$) can be ruled out a $3\sigma$ limit. The most probable solution corresponds to $P_{\rm e}=46.768131\pm0.000076\,{\rm d}$ and, assuming circular orbits, yields a relatively low-significance $2.6\sigma$ RV detection of planet e with a mass of $M_{\rm e}=0.66\pm0.26\,M_{\rm Jup}$. In the absence of additional constraints on planet e's orbital period (i.e., considering the posteriors derived for the seven most probable $P_{\rm e}$ values), we obtain a $2\sigma$ upper limit of $M_{\rm e}<1.34\,M_{\rm Jup}$.

The mass posteriors derived for planets b, c, and d are approximately independent of the assumed $P_{\rm e}$ (small correlations are apparent; see Fig. \ref{fig:K_pdf}). We obtain an $\approx2\sigma$ detection of planet c with a mass of $M_{\rm c}=19.8_{-8.9}^{+9.3}\,M_\oplus$. For planets b and d, we obtain $2\sigma$ upper mass limits of $M_{\rm b}<159\,M_\oplus$ and $M_{\rm d}<36\,M_\oplus$, respectively. We note that the injection-recovery tests that were carried out (Sect. \ref{sect:injection} of the Appendix) suggest that our model systematically underestimates the mass of planet d by $10-15\%$, which, when taken into account, increases planet d's upper mass limit to $M_{\rm d}<41.4\,M_\oplus$.

The mass constraints derived here are lower than those reported by \citet{suarezmascareno2021}. These authors obtain detections of planets b and e and derive masses of $M_{\rm b}=203\pm60\,M_\oplus$ and $M_{\rm e}=1.16\pm0.30\,M_{\rm Jup}$. They also find $2\sigma$ upper limits for planets c and d of $M_{\rm c}<76\,M_\oplus$ and $M_{\rm d}<99\,M_\oplus$. The differences between these values and those derived in our study can potentially be attributed to various factors like the inclusion of additional RV measurements, the inclusion of new \emph{TESS} observations that provide a greater constraint on planet e's orbital period, and the sensitivity of the results to the adopted stellar activity model. The GP-based approach used in our analysis and adopted by \citet{suarezmascareno2021} suggests that the accuracy with which the system's planetary RV signals can be recovered is sensitive to the choice of covariance function. Similar to the analysis carried out by \citet{benatti2021} for the young DS~Tuc~A system, we find that adopting the Quasi-Periodic kernel (Eqn. \ref{eqn:GP_QP}) yielded a higher accuracy with fewer systematic biases compared to the SHO kernel (Eqn. \ref{eqn:GP_SHO}), as evaluated using injection-recovery tests.

Most of the analysis presented in this work was carried out assuming circular orbits. When allowing for non-circular orbits for the adopted $n_{\rm e}=51$ solution, we obtained $2\sigma$ upper limits on the eccentricities of planets b, d, and e of $e_{\rm b}<0.13$, $e_{\rm d}<0.14$, and $e_{\rm e}<0.32$. We note that \citet{arevalo2022} derive a similar upper limit for planet b's eccentricity of $<0.17$ using dynamical stability constraints based on the masses reported by \citet{suarezmascareno2021}. In the case of planet c, we obtained a high eccentricity of $e_{\rm c}=0.44_{-0.12}^{+0.10}$ and a much higher mass of $M_{\rm c}=39\pm11\,M_\oplus$. However, considering (1) planet c's relatively small RV semi-amplitude ($K_{\rm c}=5.7\pm2.6\,{\rm m/s}$ assuming circular orbits), (2) the typical RV measurement uncertainty ($\approx10\,{\rm m/s}$), and (3) the fact that sparse, low signal-to-noise RV data sets are easily biased towards higher eccentricities \citep{shen2008}, we conclude that the high $e_{\rm c}$ and high $M_{\rm c}$ solution is likely biased and therefore not reliable.

\subsection{Interior Structure and Evolution}
\begin{figure}
	\centering
	\includegraphics[width=1\columnwidth]{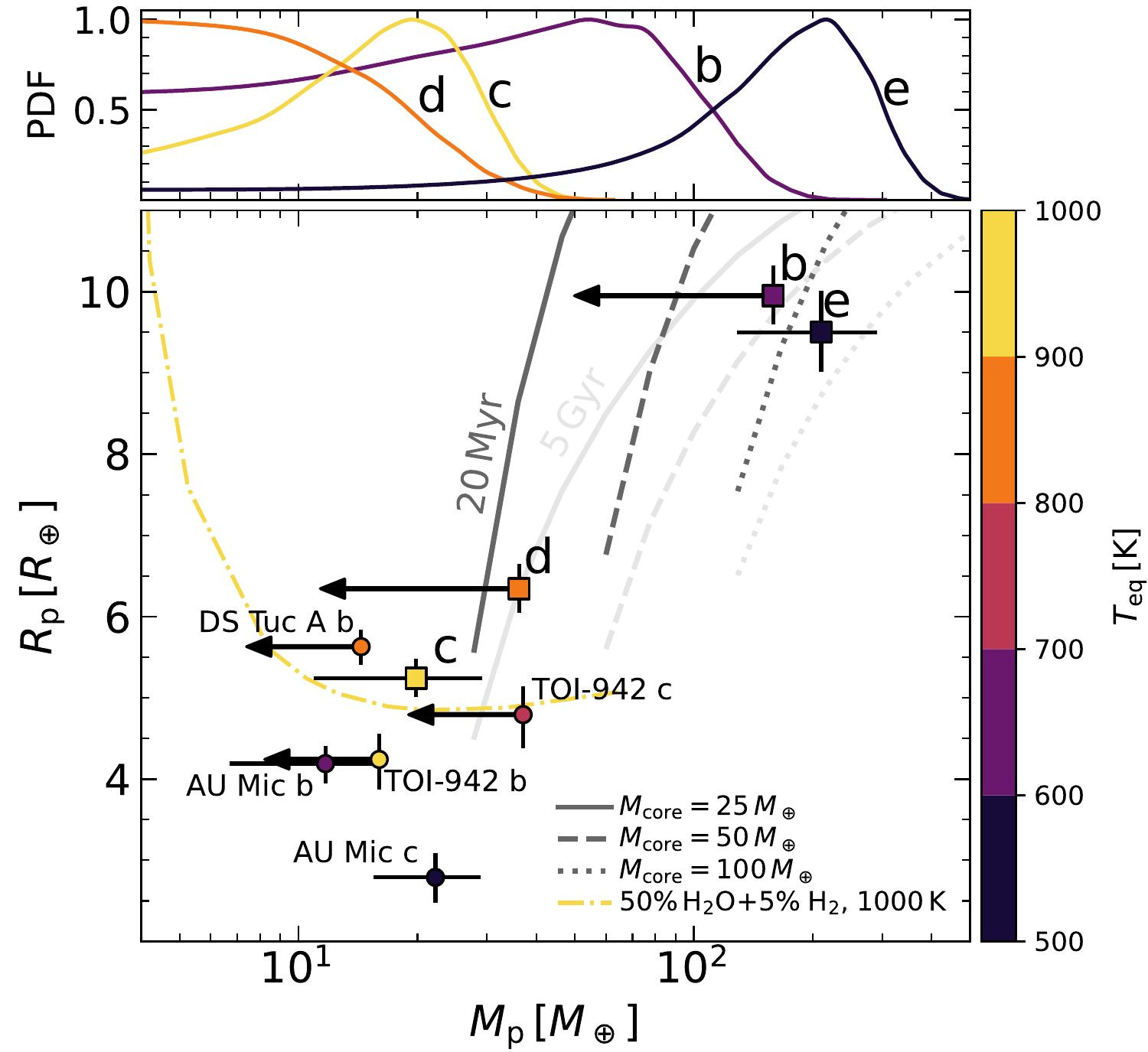}\vspace{-0.1cm}
	\caption{Masses and radii of V1298~Tau's four transiting planets derived for the most probable $P_{\rm e}$ and assuming circular orbits (squares; upper mass limits correspond to $2\sigma$) where colors correspond to equilibrium temperature. Five planets with mass constraints from three other young systems ($<100\,{\rm Myr}$) are plotted for comparison (circles): DS~Tuc~A~b \citep{benatti2021}, AU~Mic~b and c \citep{gilbert2022b,zicher2022}, and TOI-942~b and c \citep{carleo2021a}. Dark gray and light gray lines are theoretical planet mass-radius relations published by \citet{fortney2007} at an age of $20\,{\rm Myr}$ and $5\,{\rm Gyr}$, respectively. These models are computed for 50/50 ice-rock cores with H/He envelopes that are irradiated by a Sun-like host star at a distance of $0.1\,{\rm au}$. The dash-dotted curve is a model published by \citet{zeng2019} composed of an Earth-like rocky core ($47.5\%$ by mass) with an H$_2$O layer ($47.5\%$) and an H$_2$ envelope ($5\%$) at $T_{\rm eq}=1000\,{\rm K}$.}
	\label{fig:mass_radius}
\end{figure}

In Fig. \ref{fig:mass_radius}, we compare our updated mass constraints and precise radii for V1298~Tau's four transiting planets --- derived for the $n_{\rm e}=51$ ($P_{\rm e}=46.768131\pm0.000076\,{\rm d}$) solution assuming circular orbits --- with theoretical mass-radius relationships published by \citet{fortney2007}. We plot models calculated for an age of $20\,{\rm Myr}$ (black lines) and $5\,{\rm Gyr}$ (grey lines). The models consist of a core with a 50/50 mixture of ice and rock that is enshrouded by a H/He envelope; they include the effects of irradiation from a Sun-like host star at a distance of $0.1\,{\rm au}$ (V1298~Tau b, c, d, and e have semi-major axes ranging from $0.08-0.26\,{\rm au}$). We find that planet e's mass and radius are in good agreement with the $100\,M_\oplus$ core model. Based on the derived $2\sigma$ upper mass limits, planet b is approximately consistent with the models calculated for core masses of $25-100\,M_\oplus$ while planet d is consistent with the $25\,M_\oplus$ core model. Planet c's radius falls below the computed $20\,{\rm Myr}$ old evolutionary tracks, however, it is in close agreement with the model published by \citet{zeng2019} that consists of a rocky core with an outer H$_2$O layer (50/50 by mass) at an equilibrium temperature of $1000\,{\rm K}$ (cf. planet c's $T_{\rm eq}=979\,{\rm K}$).

\citet{owen2020b} demonstrates how precise mass measurements of young, gas-rich planets orbiting close to their host stars such as V1298~Tau~c can be used to test whether it's formation is consistent with the core accretion theory or if the planet has gone through the rapid mass loss ``boil-off" phase. The derived mass of $19.8_{-8.9}^{+9.3}\,M_\oplus$ and the conservative $2\sigma$ upper limit of $<39\,M_\oplus$ are both consistent with core accretion and do not require the invocation of boil-off to be explained. However, considering the low $\approx2\sigma$ significance of planet c's recovered RV signature, additional RV measurements and/or TTV measurements are needed to confirm the derived $M_{\rm c}$ and further reduce the uncertainties.

\subsection{Implications for Mass Loss}
Based on X-ray observations of V1298~Tau, \citet{poppenhaeger2020} conclude that, depending on the assumed stellar activity and planet masses, V1298~Tau's inner three planets may currently have a relatively high atmospheric mass loss rate such that their primordial H/He envelopes are eventually stripped away entirely. \citet{maggio2022} predict that planets c and d are undergoing significant atmospheric evaporation if their masses are $\lesssim40\,M_\oplus$ and $\lesssim33\,M_\oplus$, respectively. Planet c's estimated mass of $\approx20\,M_\oplus$ is therefore indicative of strong mass loss currently taking place while planet d's $2\sigma$ upper mass limit of $M_{\rm d}<36\,M_\oplus$ ($M_{\rm d}<41.4\,M_\oplus$ when accounting for the $15\%$ bias noted above) is uninformative in terms of whether the planet is undergoing mass loss. Whether evaporation is occurring may be tested by searching for excess in-transit H/He absorption \citep[e.g.,][]{oklopcic2018,allart2019,feinstein2021,vissapragada2021}. Coupled with improved mass constraints (and better period constraints for planet e), such detections would help constrain atmospheric mass loss models \citep[e.g.,][]{salz2016a,linssen2022}.

\subsection{Future RV Work}
Additional high-precision RV measurements may be able to further improve the mass constraints derived in this work. We estimated how the results derived here could be improved if an additional 120 nightly RV measurements with uncertainties of $10\,{\rm m/s}$ are included in the analysis. The simulated measurements were generated using the general injection testing framework described in Sect. \ref{sect:injection} of the Appendix. The stellar activity was estimated from the HARPS-N GP activity model calculated using the median solution listed in Table \ref{tbl:pl} and shifted to the time stamps of the simulated measurements, which were arbitrarily set to start shortly after the last MAROON-X measurement. We assumed circular orbits and planet masses of $M_{\rm b}=60\,M_\oplus$, $M_{\rm c}=20\,M_\oplus$, $M_{\rm d}=20\,M_\oplus$, and $M_{\rm e}=200\,M_\oplus$. White noise defined by the measurement uncertainties and a $5\,{\rm m/s}$ instrumental jitter was then added to each simulated RV measurement and the NUTS sampling analysis was used to estimate the resulting uncertainties. We find that including the additional simulated RV measurements yields high-significance RV detections of planets b, c, and e with mass uncertainties of $\approx5-30\,M_\oplus$.

The mass constraints derived in our analysis have relatively large uncertainties primarily due to the impact of stellar activity: we find that applying our model to a simulated data set that includes only white noise due to measurement uncertainties and instrumental jitter yields mass uncertainties that are $\lesssim10\,M_\oplus$ for all four planets. The systematic bias that causes planet d's mass to be underestimated by $\approx10-15\%$ can also be attributed to imperfect modelling of the stellar activity. Therefore, in addition to obtaining more RV measurements, the derived mass constraints can likely be improved by adopting more physically-motivated GP models \citep[e.g.][]{luger2021}, incorporating additional stellar activity tracers using 2D GPs \citep[e.g.,][]{klein2021,barragan2021}, and/or accounting for correlations with wavelength \citep{cale2021}.

\acknowledgments{We thank John Livingston, Trevor David, and Erik Petigura for useful discussions that helped to improve the work presented here. We also thank the anonymous referee for their critiques and helpful suggestions. The University of Chicago group acknowledges funding for the MAROON-X project from the David and Lucile Packard Foundation, the Heising-Simons Foundation, the Gordon and Betty Moore Foundation, the Gemini Observatory, the NSF (award number 2108465), and NASA (grant number 80NSSC22K0117). ADF acknowledges support by the National Science Foundation Graduate Research Fellowship Program under Grant No. (DGE-1746045). This work was enabled by observations made from the Gemini North telescope, located within the Maunakea Science Reserve and adjacent to the summit of Maunakea. We are grateful for the privilege of observing the Universe from a place that is unique in both it's astronomical quality and it's cultural significance.}

\facilities{Gemini-North (MAROON-X), \emph{TESS} \citep{ricker2014}, \emph{K2} \citep{borucki2010a, howell2014}. The MAROON-X observations were collected under program GN-2021B-Q-103. The TESS data presented in this paper were obtained from the Mikulski Archive for Space Telescopes (MAST) at the Space Telescope Science Institute. The specific observations analyzed can be accessed via \dataset[10.17909/72nn-2166]{https://doi.org/10.17909/72nn-2166}.}

\software{\texttt{AstroPy} \citep{astropycollaboration2013,astropycollaboration2018,astropycollaboration2022}, \texttt{celerite2} \citep{foreman-mackey2017,Foreman_Mackey_2018}, \texttt{exoplanet} \citep{foreman-mackey2021}, \texttt{matplotlib} \citep{Hunter:2007}, \texttt{numpy} \citep{harris2020array}, \texttt{pymc3} \citep{salvatier2016}, \texttt{RadVel} \citep{fulton2018a}, \texttt{scipy} \citep{2020SciPy-NMeth}, \texttt{SERVAL} \citep{zechmeister2018}, \texttt{starry} \citep{luger2019}}

\input{tables/param_tbl.tex}

\section*{Appendix}
The derived transit times and O-C values are listed in Table \ref{tbl:tt}. In Fig. \ref{fig:Mp_corner} we show the planetary mass posterior distributions for the adopted $n_{\rm e}=51$ solution; a similar plot showing the mass and eccentricity posteriors for the non-circular case are shown in Fig. \ref{fig:Mp_corner_e}.

\subsection{Injection Tests}\label{sect:injection}
Injection-recovery tests for the planetary RV signals were carried by modelling simulated RV measurements that include the planetary signals and the large stellar activity signals. For the stellar activity, we used the GP models associated with the median solution listed in Table \ref{tbl:pl} and shown in Fig. \ref{fig:rv}, which do not include the planetary signals. Simulated planetary RV signals were generated assuming circular orbits for a grid of planet masses where each of the four planets was given a mass of $0\,M_\oplus$ (i.e., no signal), $50\,M_\oplus$, $100\,M_\oplus$, or $200\,M_\oplus$. The $200\,M_\oplus$ signals associated with the planets correspond to semi-amplitudes of $\approx40-60\,{\rm m/s}$. We then injected the simulated planet-induced RVs into the activity model individually (i.e., for these tests, only a single planet's RV signal is injected at a time) and added white noise with a variance defined by the measurement uncertainties and the median instrumental jitter hyperparameters ($\sigma_i^2+\sigma_{\rm jit}^2$). The NUTS sampling analysis was then applied to the simulated RV data sets along with the observed photometry in order to estimate the mass and uncertainties associated with the injected signal.

\begin{figure}
	\centering
	\includegraphics[width=0.35\textwidth]{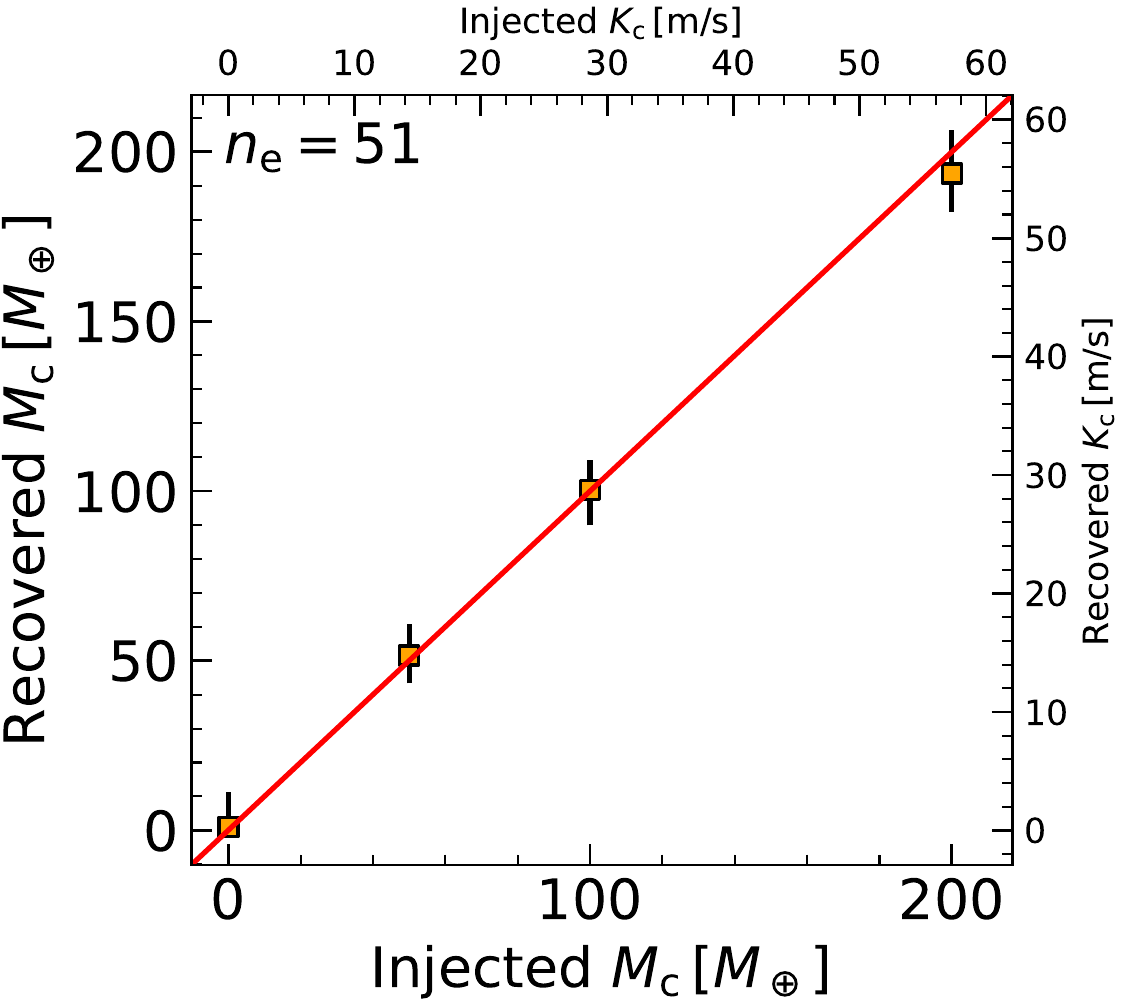}
	\includegraphics[width=0.35\textwidth]{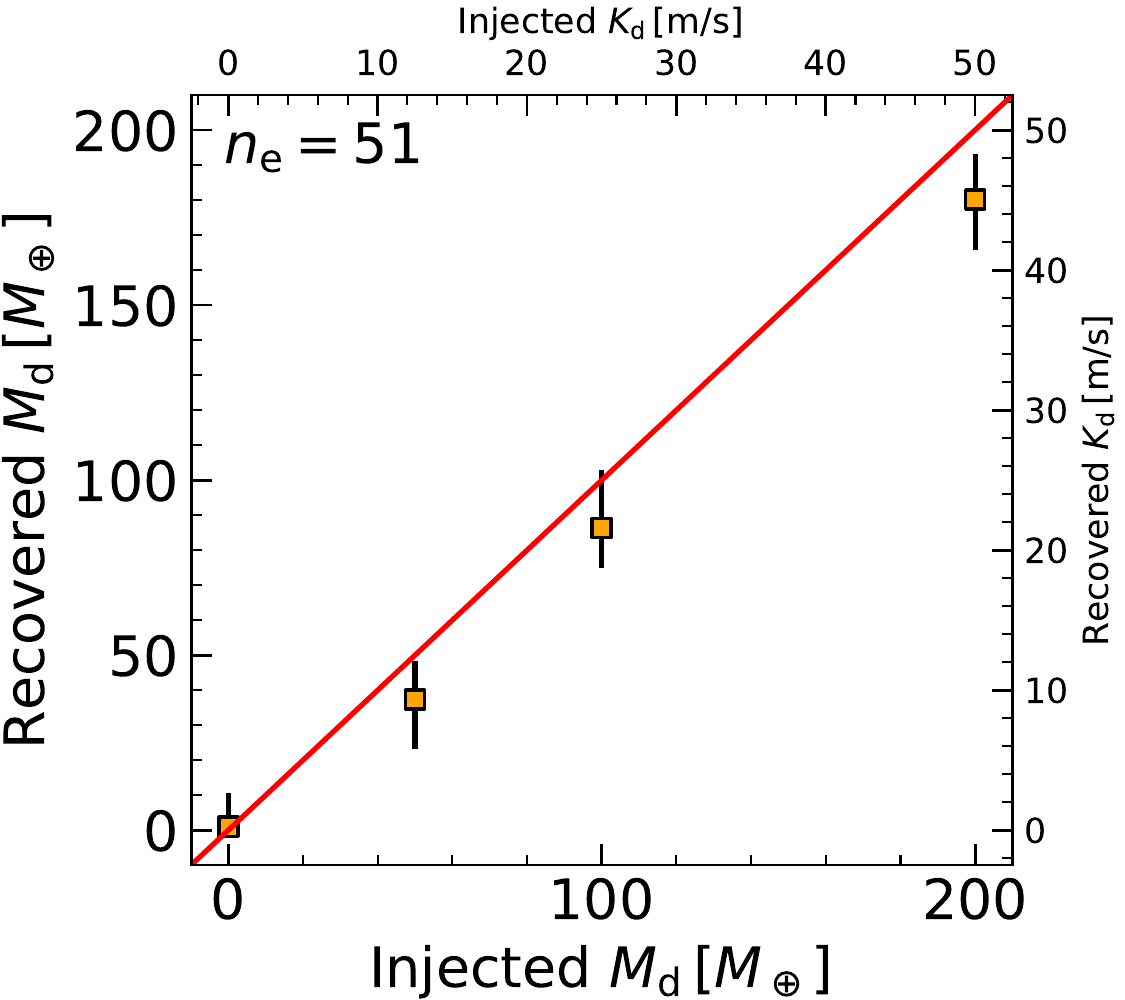}
	\includegraphics[width=0.35\textwidth]{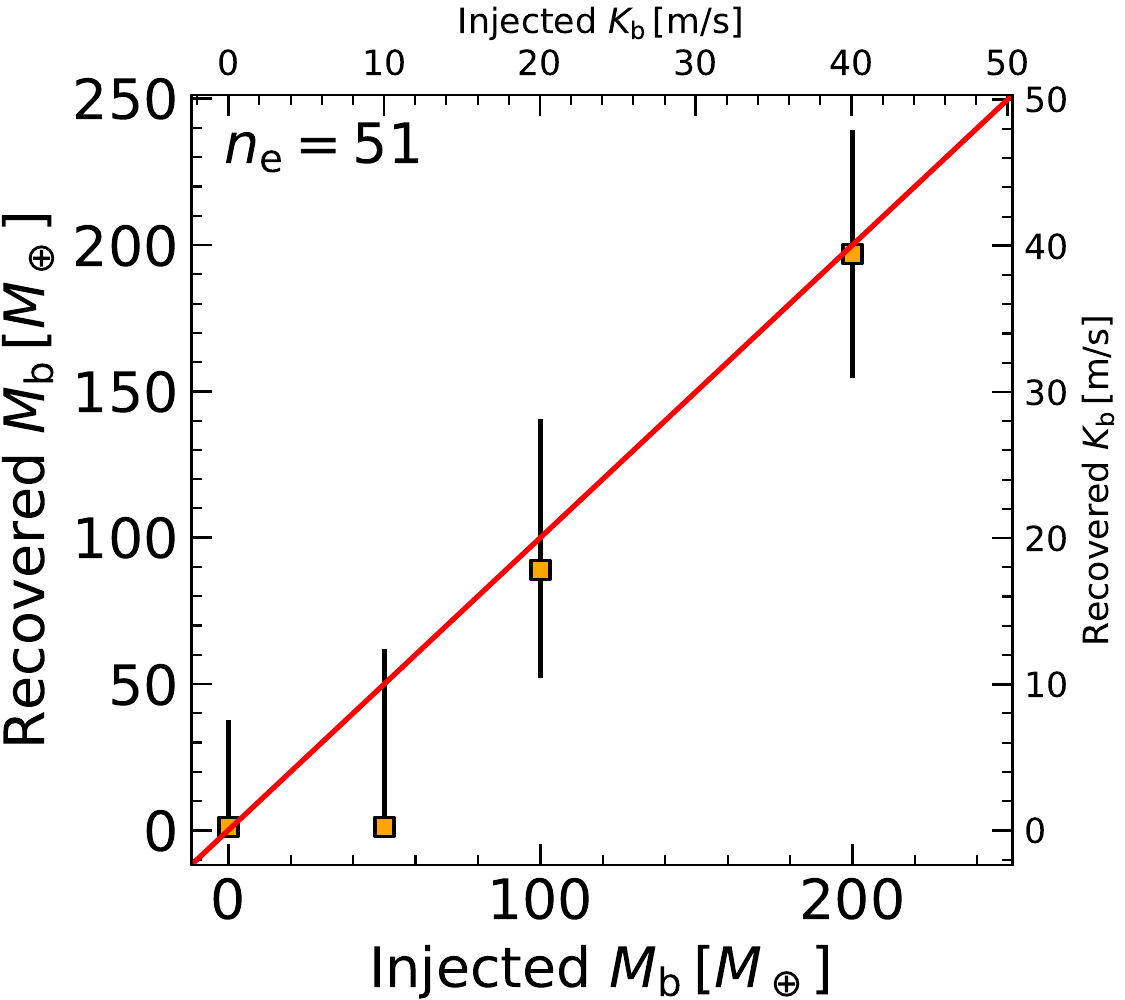}
	\includegraphics[width=0.35\textwidth]{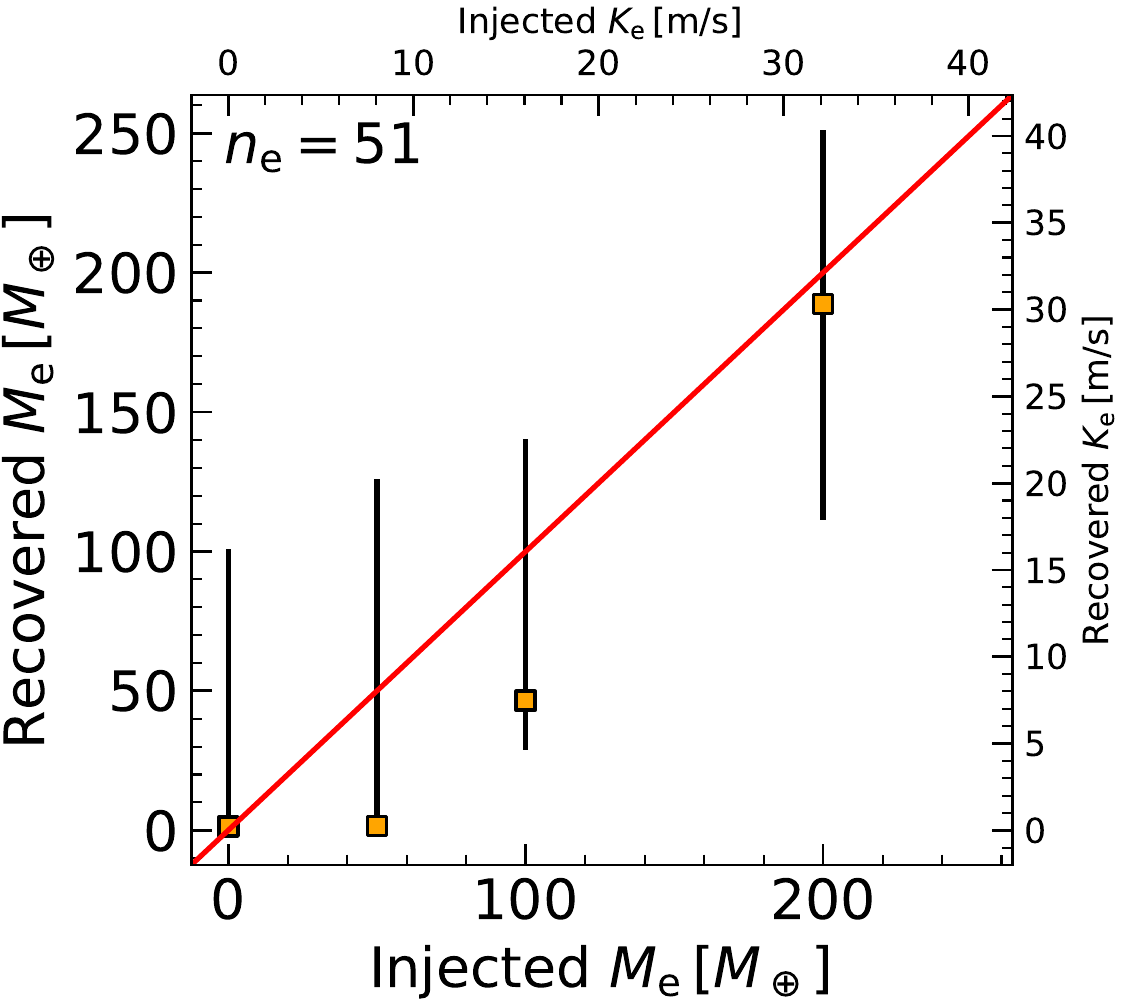}
	\caption{Injected versus recovered planetary RV signatures. Note the differing axis scales.}
	\label{fig:injection_test}
\end{figure}

\begin{figure*}
	\centering
	\includegraphics[width=2\columnwidth]{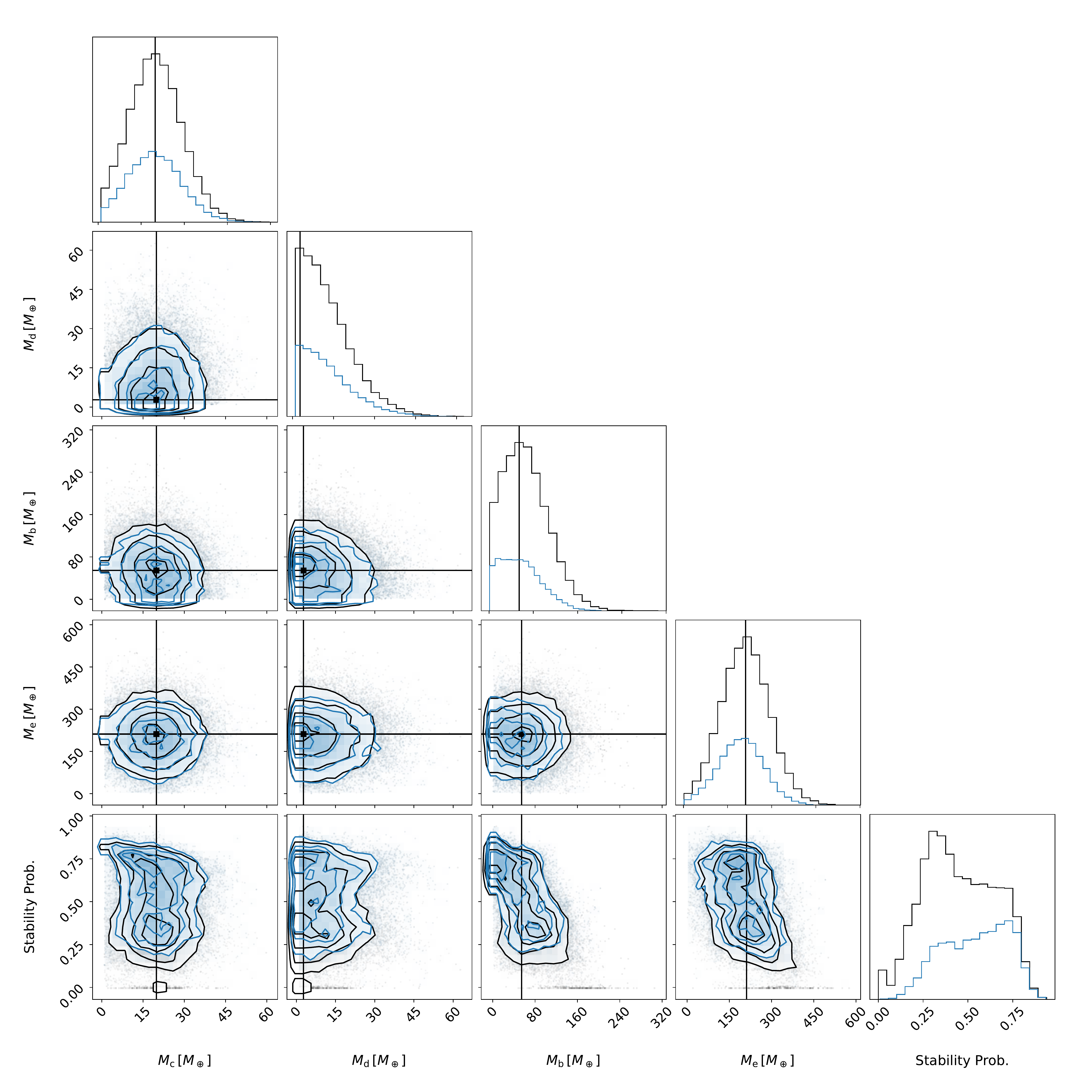}\vspace{-0.1cm}
	\caption{Marginalized planet mass posteriors derived for the most probable $n_{\rm e}=51$ solution assuming circular orbits (black contours). The dynamical stability probability ($P({\rm Stable})$) estimated for each sample using SPOCK is also shown. The blue contours show the posteriors obtained after after applying rejection sampling to the calculated $P({\rm Stable})$ values.}
	\label{fig:Mp_corner}
\end{figure*}

\begin{figure*}
	\centering
	\includegraphics[width=2\columnwidth]{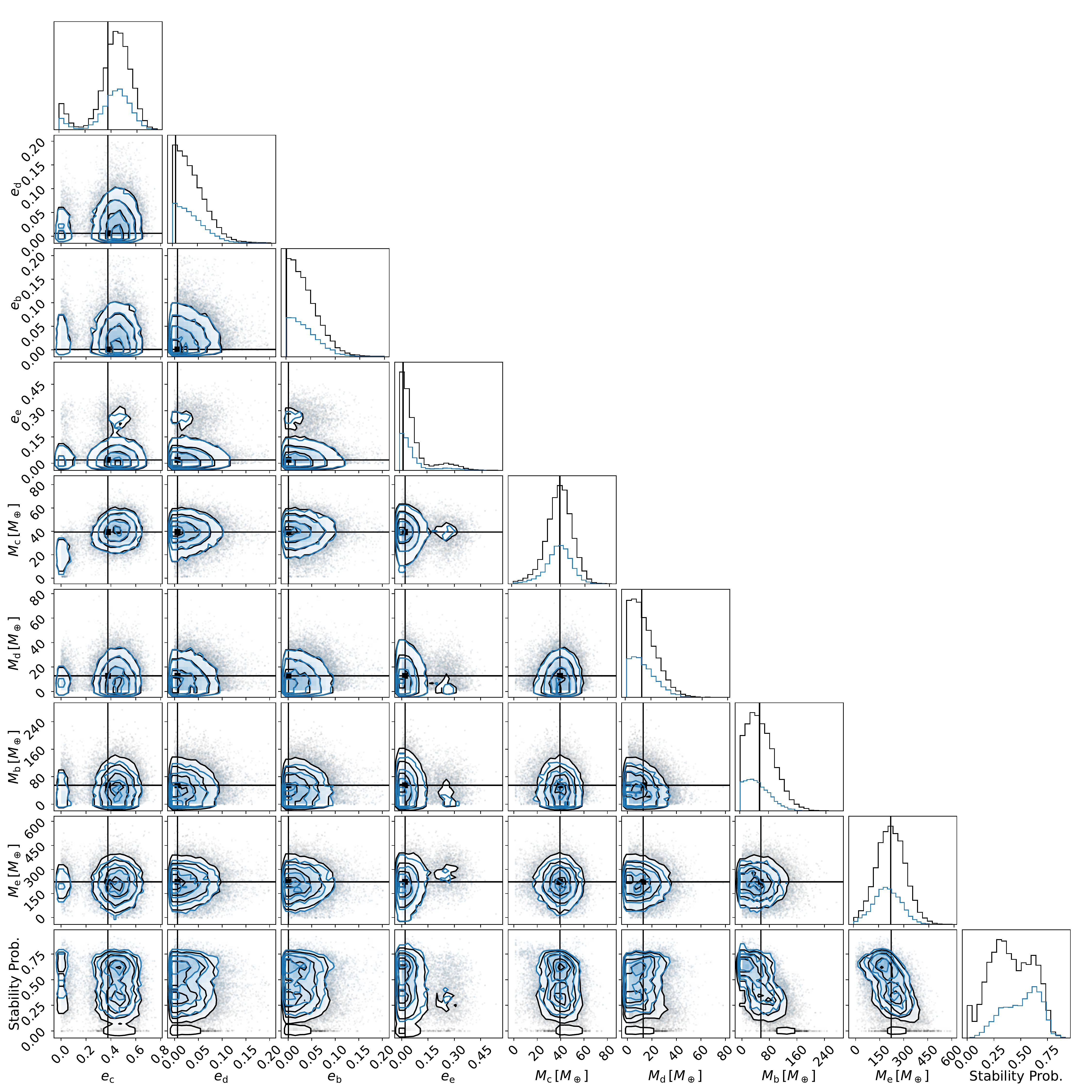}\vspace{-0.1cm}
	\caption{Same as Fig. \ref{fig:Mp_corner} but for non-circular orbits where the eccentricity distributions are also plotted.}
	\label{fig:Mp_corner_e}
\end{figure*}

The results of the injection-recovery tests are shown in Fig. \ref{fig:injection_test}. We find that planet c exhibits the highest accuracy and the smallest uncertainties with all injected signals being recovered within $0.5\sigma$. Planets b and e have significantly larger uncertainties and show that injected signals with $M_{\rm b}\lesssim50\,M_\oplus$ ($K_{\rm b}\lesssim10\,{\rm m/s}$) and $M_{\rm e}\lesssim100\,M_\oplus$ ($K_{\rm e}\lesssim15\,{\rm m/s}$) are not detected; the recovered masses all agree with the injected values with $0.5\sigma$. For planet d, a systematic bias is apparent in which the recovered masses are $\approx10-15\,\%$ lower than the masses of the injected signals (corresponding to a difference in semi-amplitude of $\approx3-5\,{\rm m/s}$). In this case, the recovered $M_{\rm d}$ for injected masses of $50-200\,M_\oplus$ are in agreement within $1.5\sigma$.

\input{tables/ttv_tbl}

\bibliography{V1298_Tau}{}
\bibliographystyle{aasjournal}

\end{document}

%% file: tables/param_tbl.tex
\begin{table*}
	\caption{Planetary, stellar, and additional model parameters derived using the $n_{\rm e}=51$ solution ($P_{\rm e}\approx46.8\,{\rm d}$). The adopted values correspond to the median, the errors are $1\sigma$ (taken to be the $15.9$ and $84.1$ percentiles), and upper/lower limits correspond to $2\sigma$ ($2.3$ and $97.7$ percentiles). The equilibrium temperatures ($T_{\rm eq}$) are calculated assuming zero albedo. Note that the mass constraints derived for planet d are likely underestimated by $10-15\%$ based on the injection-recovery tests presented in Sect. \ref{sect:injection} of the Appendix. Values presented in parentheses correspond to the uncertainty in the last digits.}\vspace{-0.6cm}
	\label{tbl:pl}
	\begin{center}
	\begin{tabular}{@{\extracolsep{\fill}}l c c c c c r@{\extracolsep{\fill}}}
		\hline
		\hline
		\noalign{\vskip0.5mm}
Parameter & c & d & b & e & & \\
\hline
$T_0-2454833$                       & $2231.2822(22)$ & $2239.3943(17)$ & $2234.0481(12)$ & $2263.6222(27)$ \\
$P\,{\rm [d]}$                      & $8.248720(24)$ & $12.402140(17)$ & $24.140410(22)$ & $46.768131(76)$ \\
$b$                                 & $<0.36$ & $<0.35$ & $0.451_{-0.030}^{+0.033}$ & $0.595_{-0.024}^{+0.025}$ \\
$a/R_\star$                         & $13.32_{-0.24}^{+0.20}$ & $17.48_{-0.32}^{+0.26}$ & $27.25_{-0.49}^{+0.40}$ & $42.35_{-0.77}^{+0.62}$ \\
$R_{\rm p}/R_\star$                 & $0.0354\pm0.0013$ & $0.0429_{-0.0017}^{+0.0016}$ & $0.0673\pm0.0017$ & $0.0643\pm0.0029$ \\
$K\,{\rm [m/s]}$                    & $5.7\pm2.6$ & $<9.0$ & $<32$ & $34\pm13$ \\
$M_{\rm p}/M_\star\,\times10^4$     & $0.52_{-0.23}^{+0.24}$ & $<0.94$ & $<4.1$ & $5.5\pm2.1$ \\
$i\,[{\rm deg}]$                    & $>88.39$ & $>88.80$ & $89.052_{-0.087}^{+0.072}$ & $89.195_{-0.044}^{+0.040}$ \\
$a\,[{\rm au}]$                     & $0.0839\pm0.0014$ & $0.1101_{-0.0019}^{+0.0018}$ & $0.1716_{-0.0029}^{+0.0028}$ & $0.2667_{-0.0045}^{+0.0043}$ \\
$T_{\rm 14}\,[{\rm hrs}]$           & $4.826_{-0.087}^{+0.080}$ & $5.596_{-0.081}^{+0.063}$ & $6.547_{-0.068}^{+0.070}$ & $7.45_{-0.11}^{+0.12}$ \\
$T_{\rm eq}\,[{\rm K}]$             & $979\pm21$ & $855\pm19$ & $685\pm15$ & $549\pm12$ \\
$R_{\rm p}\,[R_{\rm Jup}]$          & $0.467\pm0.021$ & $0.566_{-0.026}^{+0.027}$ & $0.888_{-0.031}^{+0.033}$ & $0.848_{-0.044}^{+0.046}$ \\
$R_{\rm p}\,[R_\oplus]$             & $5.24_{-0.23}^{+0.24}$ & $6.34\pm0.30$ & $9.95_{-0.35}^{+0.37}$ & $9.50_{-0.49}^{+0.51}$ \\
$M_{\rm p}\,[M_{\rm Jup}]$          & $0.062_{-0.028}^{+0.029}$ & $<0.114$ & $<0.50$ & $0.66\pm0.26$ \\
$M_{\rm p}\,[M_\oplus]$             & $19.8_{-8.9}^{+9.3}$ & $<36$ & $<159$ & $210\pm82$ \\
$\rho_{\rm p}\,[{\rm g/cm^3}]$      & $0.76_{-0.35}^{+0.38}$ & $<0.81$ & $<0.90$ & $1.33_{-0.53}^{+0.59}$ \\

\hline
Parameter & & & & & & \\
\hline
$M_\star\,[M_\odot]$                & $1.157_{-0.058}^{+0.057}$ & & & & & \\
$R_\star\,[R_\odot]$                & $1.355_{-0.030}^{+0.032}$ & & & & & \\
$T_{\rm eff}\,[{\rm K}]$            & $5050\pm100$ & & & & & \\

\hline
Parameter & \emph{K2} & \emph{TESS} & $LCOGT$ & & & \\
\hline
$u_1$                               & $0.40\pm0.18$ & $0.46_{-0.26}^{+0.23}$ & & & & \\
$u_2$                               & $<0.60$ & $0.29_{-0.27}^{+0.24}$ & & & & \\
$\langle f\rangle\,[{\rm ppt}]$     & $0.02\pm0.31$ & $0.12\pm0.61$ & $1.48\pm0.54$ & & & \\
$\ln P_{\rm rot}/[{\rm d}]$         & $1.0541\pm0.0082$ & $1.058_{-0.031}^{+0.027}$ & $1.0682\pm0.0012$ & & & \\
$\ln A/{\rm [ppt]}$                 & $-2.44_{-0.31}^{+0.40}$ & $-3.85\pm0.34$ & $0.19_{-0.75}^{+0.94}$ & & & \\
$\ln f_{\rm mix}$                   & $0.50_{-0.53}^{+0.67}$ & $0.61_{-0.47}^{+0.48}$ & $-2.95_{-0.91}^{+0.79}$ & & & \\
$\ln Q_0$                           & $3.05_{-0.52}^{+0.80}$ & $0.59_{-0.27}^{+0.30}$ & $6.25_{-0.82}^{+0.97}$ & & & \\
$\ln\sigma_{\rm Jit}/{\rm [ppt]}$   & & & $1.92_{-0.11}^{+0.08}$ & & & \\

\hline
Parameter & HARPS-N & CARMENES & STELLA & HERMES & MX (blue) & MX (red) \\
\hline
$\gamma_0$                         & $-2.1_{-10.0}^{+9.9}$ & $-0.8_{-9.8}^{+9.7}$ & $-1\pm10$ & $-0.6_{-9.8}^{+9.9}$ & $-0.6_{-9.9}^{+9.8}$ & $-1.1_{-9.5}^{+9.8}$ \\
$\ln A_1/[{\rm m/s}]$              & $5.41_{-0.10}^{+0.11}$ & $5.43_{-0.16}^{+0.15}$ & $5.63\pm0.16$ & $5.95_{-0.18}^{+0.19}$ & $5.31_{-0.22}^{+0.23}$ & $4.72_{-0.16}^{+0.18}$ \\
$\ln P_{\rm rot}/[{\rm d}]$        & $1.0636_{-0.0015}^{+0.0014}$ & & & & $1.0728_{-0.0076}^{+0.0079}$ & \\
$\ln\lambda_{\rm e}$               & $3.17_{-0.10}^{+0.09}$ & & & &  $1.94_{-0.13}^{+0.17}$ & \\
$\ln\lambda_{\rm p}$               & $-1.171_{-0.079}^{+0.082}$ & & & & $-1.34_{-0.29}^{+0.24}$ & \\
$\ln \sigma_{\rm Jit}/[{\rm m/s}]$ & $2.79_{-0.18}^{+0.16}$ & $1.4_{-2.4}^{+1.7}$ & $2.1_{-2.0}^{+1.3}$ & $3.8_{-1.7}^{+0.5}$ & $1.7_{-2.7}^{+1.1}$ & $1.7_{-2.6}^{+1.0}$ \\
\hline
Parameter & Prior & Parameter & Prior & Parameter & Prior & \\
\hline
$\ln T_0/{\rm d}$ & $\mathcal{N}(\ln T_0^\dagger,0.01)$                       & $R_\star/R_\odot$ & $\mathcal{N}(1.278,0.07)^\ddagger$               & $\ln A$ & $\mathcal{U}[-10,10]$ \\
$\ln P/{\rm d}$ & $\mathcal{N}(\ln P^\dagger,0.01)$                           & $q_1^{\dagger\dagger}$ & $\mathcal{U}[0,1]$                          & $\ln f_{\rm mix}$ & $\mathcal{U}[-5,5]$ \\
$b$ & $\mathcal{U}[0,1+R_{\rm p}/R_\star]$                                    & $q_2^{\dagger\dagger}$ & $\mathcal{U}[0,1]$                          & $\ln Q_0$ & $\mathcal{N}(1,10)$ \\
$\ln{R_{\rm p}/R_\star}$ & $\mathcal{N}(\ln R_{\rm p}/R_\star,0.1)^\dagger$   & $\langle f\rangle/{\rm ppt}$ & $\mathcal{N}(0,10)$                   & $\gamma_0/[{\rm m/s}]$ & $\mathcal{N}(0,10)$  \\
$\ln{M_{\rm p}/M_\oplus}$ & $\mathcal{U}[0,7.37]$                             & $\ln \sigma_{\rm Jit}$ & $\mathcal{N}(\ln\langle\sigma_i\rangle,3)$ & $\ln\lambda_{\rm e}$ & $\mathcal{U}[0,10]$ \\
$M_\star/M_\odot$ & $\mathcal{N}(1.17,0.06)^\ddagger$                         & $\ln P_{\rm rot}/{\rm d}$ & $\mathcal{U}[1.0,1.1]$                   & $\ln\lambda_{\rm p}$ & $\mathcal{U}[-5,1]$ \\
\hline
		\noalign{\vskip0.5mm}
		\hline
\multicolumn{7}{l}{$^\dagger$TTV analysis where $(T_{0,b},P_{\rm b})=(2234.048,24.140)$, $(T_{0,c},P_{\rm c})=(2231.280,8.249)$, $(T_{0,d},P_{\rm d})=(2239.396,12.402)$,} \\	
\multicolumn{7}{l}{$(T_{0,e},P_{\rm e})=(2263.620,43.367)$.} \\
\multicolumn{7}{l}{$^\ddagger$\citet{suarezmascareno2021}. $^{\dagger\dagger}$Parameterization of $u_1$ and $u_2$ from \citet{kipping2013}. $^*$\citet{david2019}.}\\
	\end{tabular}
	\end{center}
\end{table*}

%% file: tables/ttv_tbl.tex
\begin{table}
       \caption{Transit times (BJD$_{\rm TT}$) and O$-$C times derived from the \emph{K2} and \emph{TESS} light curves. Gaussian priors on BJD$_{\rm TT}$ -- centered on the transit times associated with the ephemerides published by \citet{david2019} and \citet{feinstein2022} -- with standard deviations of $0.05\,{\rm d}$ were adopted.}{\vspace{0cm}}
       \label{tbl:tt}
       \begin{tabular}{@{\extracolsep{\fill}}l c r@{\extracolsep{\fill}}}
              \hline
              \hline
              \noalign{\vskip0.5mm}
Planet & BJD$_{\rm TT}$ & $O-C$ \\
       & $-2454833$ & [min] \\
\hline

b & $2234.0483_{-0.0019}^{+0.0016}$ & $0.7_{-2.6}^{+2.5}$ \\
  & $2258.1897_{-0.0022}^{+0.0025}$ & $2.4_{-2.8}^{+3.0}$ \\
  & $^*2282.3265_{-0.0034}^{+0.0029}$ & $-3.1_{-3.6}^{+3.2}$ \\
  & $^*4648.0893_{-0.0029}^{+0.0028}$ & $0.2\pm2.7$ \\
  & $4672.2295\pm0.0025$ & $-0.1\pm2.7$ \\
\hline

c & $2231.2786_{-0.0077}^{+0.0067}$ & $-4_{-11}^{+10}$ \\
  & $^*2239.5287_{-0.0052}^{+0.0064}$ & $-2_{-9}^{+10}$ \\
  & $2247.7775_{-0.0066}^{+0.0086}$ & $-1_{-9}^{+12}$ \\
  & $2256.0339_{-0.0066}^{+0.0070}$ & $9_{-9}^{+10}$ \\
  & $^*2264.2536_{-0.0086}^{+0.0056}$ & $-33_{-10}^{+7}$ \\
  & $2272.5285_{-0.0074}^{+0.0070}$ & $5\pm10$ \\
  & $2280.794_{-0.006}^{+0.011}$ & $31_{-9}^{+13}$ \\
  & $^*2289.0184_{-0.0053}^{+0.0049}$ & $-5.8_{-6.8}^{+6.5}$ \\
  & $2297.2700_{-0.0068}^{+0.0057}$ & $-1.9_{-8.9}^{+8.0}$ \\
  & $^*4648.1627_{-0.0084}^{+0.0067}$ & $7_{-16}^{+14}$ \\
  & $4656.399_{-0.013}^{+0.031}$ & $-6_{-22}^{+34}$ \\
  & $4664.652_{-0.014}^{+0.023}$ & $-2_{-21}^{+28}$ \\
  & $4672.897_{-0.011}^{+0.010}$ & $-11_{-20}^{+16}$ \\
  & $4681.156_{-0.025}^{+0.028}$ & $6_{-31}^{+34}$ \\
  & $4689.398_{-0.010}^{+0.016}$ & $-3_{-17}^{+21}$ \\
\hline

d & $^*2239.3925_{-0.0036}^{+0.0034}$ & $-6.3_{-5.2}^{+5.0}$ \\
  & $2251.7922_{-0.0042}^{+0.0034}$ & $-9.9_{-5.7}^{+5.0}$ \\
  & $^*2264.2002_{-0.0058}^{+0.0087}$ & $-1_{-7}^{+10}$ \\
  & $2276.5974_{-0.0042}^{+0.0045}$ & $-8.1_{-5.7}^{+5.9}$ \\
  & $^*2289.0224_{-0.0034}^{+0.0032}$ & $24.6_{-5.0}^{+4.8}$ \\
  & $4645.4104_{-0.0067}^{+0.0050}$ & $2.5_{-8.3}^{+6.8}$ \\
  & $4657.8161_{-0.0044}^{+0.0066}$ & $8.6_{-6.1}^{+7.9}$ \\
  & $4670.2136_{-0.0045}^{+0.0055}$ & $1.6_{-6.3}^{+7.1}$ \\
  & $4682.6052_{-0.0046}^{+0.0048}$ & $-13.6_{-6.5}^{+6.4}$ \\
\hline

e & $2263.6224_{-0.0027}^{+0.0028}$ & $0$ \\
  & $4648.7969\pm0.0029$ & $0$ \\
              \noalign{\vskip0.5mm}
              \hline
\multicolumn{3}{l}{$^*$Overlapping transits or partial event coverage.} \\
       \end{tabular}
\end{table}